\documentclass[twocolumn]{aastex631}
\usepackage[utf8]{inputenc}
\usepackage{makecell}
\usepackage{amsmath}
\usepackage{tikz}
\usepackage{hyperref}
\usepackage{soul}

\newcommand{\revise}[1]{#1}
\setstcolor{red}
\newcommand{\vmax}{v_\mathrm{rot}}
\newcommand{\voversigma}{v_\mathrm{rot}/\sigma_0}
\newcommand{\fgas}{f_\mathrm{gas}}
\newcommand{\logOH}{12+\log(\mathrm{O/H})}
\newcommand{\withinre}{r<r_\mathrm{e}}

\received{---}
\revised{---}
\accepted{---}

\shorttitle{
Statistics on the Kinematics of EMPGs
}
\shortauthors{Xu et al.}

\begin{document}
\title{
EMPRESS. XII. \\
Statistics on the Dynamics and Gas Mass Fraction of 
Extremely Metal-Poor Galaxies
%Kinematics of Extremely-Metal Poor Galaxies
%and Scaling Relations Towards Primordial Systems
}

\author[0000-0002-5768-8235]{Yi Xu}
\affiliation{Institute for Cosmic Ray Research, The University of Tokyo, 5-1-5 Kashiwanoha, Kashiwa, Chiba 277-8582, Japan}
\affiliation{Department of Astronomy, Graduate School of Science, the University of Tokyo, 7-3-1 Hongo, Bunkyo, Tokyo 113-0033, Japan}
\correspondingauthor{Yi Xu}
\email{xuyi@icrr.u-tokyo.ac.jp}

\author[0000-0002-1049-6658]{Masami Ouchi}
\affiliation{National Astronomical Observatory of Japan, 2-21-1 Osawa, Mitaka, Tokyo 181-8588, Japan}
\affiliation{Institute for Cosmic Ray Research, The University of Tokyo, 5-1-5 Kashiwanoha, Kashiwa, Chiba 277-8582, Japan}
\affiliation{Kavli Institute for the Physics and Mathematics of the Universe (WPI), University of Tokyo, Kashiwa, Chiba 277-8583, Japan}

\author[0000-0001-7730-8634]{Yuki Isobe}
\affiliation{Institute for Cosmic Ray Research, The University of Tokyo, 5-1-5 Kashiwanoha, Kashiwa, Chiba 277-8582, Japan}
\affiliation{Department of Physics, Graduate School of Science, The University of Tokyo, 7-3-1 Hongo, Bunkyo, Tokyo 113-0033, Japan}

\author[0000-0003-2965-5070]{Kimihiko Nakajima}
\affiliation{National Astronomical Observatory of Japan, 2-21-1 Osawa, Mitaka, Tokyo 181-8588, Japan}

\author[0000-0002-5443-0300]{Shinobu Ozaki}
\affiliation{National Astronomical Observatory of Japan, 2-21-1 Osawa, Mitaka, Tokyo 181-8588, Japan}

\author[0000-0003-0068-9920]{Nicolas F. Bouch{\'e}}
\affiliation{Univ Lyon, Univ Lyon1, ENS de Lyon, CNRS, Centre de Recherche Astrophysique de Lyon UMR5574, F-69230 Saint-Genis-Laval France}

\author[0000-0003-1173-8847]{John H. Wise}
\affiliation{Center for Relativistic Astrophysics, School of Physics, Georgia Institute of Technology, Atlanta, GA 30332, USA}

\author[0000-0002-6155-7166]{Eric Emsellem}
\affiliation{European Southern Observatory, Karl-Schwarzschild-Stra{\ss}e 2, 85748 Garching, Germany}
\affiliation{Univ Lyon, Univ Lyon1, ENS de Lyon, CNRS, Centre de Recherche Astrophysique de Lyon UMR5574, F-69230 Saint-Genis-Laval France}

\author[0000-0002-3801-434X]{Haruka Kusakabe} 
\affiliation{Observatoire de Gen{\'e}ve, Universit{\'e} de Gen{\'e}ve, 51 Ch. des Maillettes, 1290 Versoix, Switzerland}

\author[0000-0002-8996-7562]{Takashi Hattori}
%\affiliation{Subaru Telescope, National Astronomical Observatory of Japan, 650 North A’ohoku Place, Hilo, HI 96720, USA}
\affiliation{Subaru Telescope, National Astronomical Observatory of Japan, National Institutes of Natural Sciences (NINS), 650 North A'ohoku Place, Hilo, HI 96720, USA}

\author[0000-0002-7402-5441]{Tohru Nagao}
\affiliation{Research Center for Space and Cosmic Evolution, Ehime
University, Bunkyo-cho 2-5, Matsuyama, Ehime 790-8577, Japan}

\author[0000-0001-6246-2866]{Gen Chiaki}
\affiliation{National Astronomical Observatory of Japan, 2-21-1 Osawa, Mitaka, Tokyo 181-8588, Japan}

\author[0000-0002-0547-3208]{Hajime Fukushima}
\affiliation{Center for Computational Sciences, University of Tsukuba, Ten-nodai, 1-1-1 Tsukuba, Ibaraki 305-8577, Japan}

\author[0000-0002-6047-430X]{Yuichi Harikane} 
\affiliation{Institute for Cosmic Ray Research, The University of Tokyo, 5-1-5 Kashiwanoha, Kashiwa, Chiba 277-8582, Japan}
\affiliation{Department of Physics and Astronomy, University College London, Gower Street, London WC1E 6BT, UK}

\author[0000-0002-8758-8139]{Kohei Hayashi}
\affiliation{National Institute of Technology, Ichinoseki College, Hagisho, Ichinoseki, 021-8511, Japan}
\affiliation{Astronomical Institute, Tohoku University, 6-3 Aoba, Aramaki, Aoba-ku, Sendai, Miyagi 980-8578, Japan}
\affiliation{Institute for Cosmic Ray Research, The University of Tokyo, 5-1-5 Kashiwanoha, Kashiwa, Chiba 277-8582, Japan}

\author[0000-0002-5661-033X]{Yutaka Hirai}
% \altaffiliation{JSPS Research Fellow}
\affiliation{Department of Physics and Astronomy, University of Notre Dame, 225 Nieuwland Science Hall, Notre Dame, IN 46556, USA}
\affiliation{Astronomical Institute, Tohoku University, 6-3 Aoba, Aramaki, Aoba-ku, Sendai, Miyagi 980-8578, Japan}

\author[0000-0002-1418-3309]{Ji Hoon Kim}
\affiliation{Astronomy Program, Department of Physics and Astronomy, Seoul National University, 1 Gwanak-ro, Gwanak-gu, Seoul 08826, Republic of Korea}
\affiliation{SNU Astronomy Research Center, Seoul National University, 1 Gwanak-ro, Gwanak-gu, Seoul 08826, Republic of Korea}

\author[0000-0003-0695-4414]{Michael V. Maseda}
\affiliation{Department of Astronomy, University of Wisconsin-Madison, 475 N. Charter Street, Madison, WI 53706, USA}

\author[0000-0001-7457-8487]{Kentaro Nagamine}
\affiliation{Theoretical Astrophysics, Department of Earth \& Space Science, Graduate School of Science, Osaka University, 
1-1 Machikaneyama, Toyonaka, Osaka 560-0043, Japan}
%\affiliation{Kavli IPMU (WPI), The University of Tokyo, 5-1-5 Kashiwanoha, Kashiwa, Chiba 277-8583, Japan}
\affiliation{Kavli Institute for the Physics and Mathematics of the Universe (WPI), University of Tokyo, Kashiwa, Chiba 277-8583, Japan}
\affiliation{Department of Physics \& Astronomy, University of Nevada, Las Vegas, 4505 S. Maryland Pkwy, Las Vegas, NV 89154-4002, USA}

\author{Takatoshi Shibuya}
\affiliation{Kitami Institute of Technology, 165 Koen-cho, Kitami, Hokkaido 090-8507, Japan}

\author[0000-0001-6958-7856]{Yuma Sugahara} 
\affiliation{National Astronomical Observatory of Japan, 2-21-1 Osawa, Mitaka, Tokyo 181-8588, Japan}
\affiliation{Waseda Research Institute for Science and Engineering, Faculty of Science and Engineering, Waseda University, 3-4-1, Okubo, Shinjuku, Tokyo 169-8555, Japan}

\author[0000-0002-1319-3433]{Hidenobu Yajima}
\affiliation{Center for Computational Sciences, University of Tsukuba, Ten-nodai, 1-1-1 Tsukuba, Ibaraki 305-8577, Japan}

%%%%%%%%%%%%%%%%%%%%%%%%%%%%%
%Shohei Aoyama
\author[0000-0002-1005-4120]{Shohei Aoyama}
\affiliation{Institute of Management and Information Technologies, Chiba University, 1-33, Yayoi-cho, Inage-ward, Chiba, 263-8522, Japan}
\affiliation{Institute for Cosmic Ray Research, The University of Tokyo, 5-1-5 Kashiwanoha, Kashiwa, Chiba 277-8582, Japan}

%Seiji Fujimoto
\author[0000-0001-7201-5066]{Seiji Fujimoto} 
\affiliation{Cosmic DAWN Center}
\affiliation{Niels Bohr Institute, University of Copenhagen, Lyngbyvej2, DK-2100, Copenhagen, Denmark}
\affiliation{Research Institute for Science and Engineering, Waseda University, 3-4-1 Okubo, Shinjuku, Tokyo 169-8555, Japan}
\affiliation{National Astronomical Observatory of Japan, 2-21-1 Osawa, Mitaka, Tokyo 181-8588, Japan}
\affiliation{Institute for Cosmic Ray Research, The University of Tokyo, 5-1-5 Kashiwanoha, Kashiwa, Chiba 277-8582, Japan}

%Keita Fukushima
\author{Keita Fukushima}
\affiliation{Theoretical Astrophysics, Department of Earth \& Space Science, Graduate School of Science, Osaka University, 
1-1 Machikaneyama, Toyonaka, Osaka 560-0043, Japan}

%Takuya Hashimoto
%\author[0000-0002-0898-4038]{Takuya Hashimoto}
%\affiliation{Tomonaga Center for the History of the Universe (TCHoU), Faculty of Pure and Applied Sciences, University of Tsukuba, Tsukuba, Ibaraki 305-8571, Japan}

%Shun Hatano
\author{Shun Hatano}
\affiliation{Department of Astronomical Science, SOKENDAI (The Graduate University for Advanced Studies), Osawa 2-21-1, Mitaka, Tokyo, 181-8588, Japan}

%Akio Inoue
\author[0000-0002-7779-8677]{Akio K. Inoue}
\affiliation{Waseda Research Institute for Science and Engineering, Faculty of Science and Engineering, Waseda University, 3-4-1, Okubo, Shinjuku, Tokyo 169-8555, Japan}
\affiliation{Department of Physics, School of Advanced Science and Engineering, Faculty of Science and Engineering, Waseda University, 3-4-1 Okubo, Shinjuku, Tokyo 169-8555, Japan}

%Tsuyoshi Ishigaki
\author{Tsuyoshi Ishigaki}
\affiliation{Department of Physical Science and Materials Engineering, Faculty of Science and Engineering, Iwate University \\
3-18-34 Ueda, Morioka, Iwate 020-8550, Japan}

%Masahiro Kawasaki
\author{Masahiro Kawasaki}
\affiliation{Institute for Cosmic Ray Research, The University of Tokyo, 5-1-5 Kashiwanoha, Kashiwa, Chiba 277-8582, Japan}
\affiliation{Kavli Institute for the Physics and Mathematics of the Universe (WPI), University of Tokyo, Kashiwa, Chiba 277-8583, Japan}

%Takashi Kojima
\author[0000-0001-5780-1886]{Takashi Kojima}
\affiliation{Institute for Cosmic Ray Research, The University of Tokyo, 5-1-5 Kashiwanoha, Kashiwa, Chiba 277-8582, Japan}
\affiliation{Department of Physics, Graduate School of Science, The University of Tokyo, 7-3-1 Hongo, Bunkyo, Tokyo 113-0033, Japan}

%Yutaka Komiyama
\author[0000-0002-3852-6329]{Yutaka Komiyama} 
\affiliation{Department of Advanced Sciences, Faculty of Science and Engineering, Hosei University, 3-7-2 Kajino-cho, Koganei-shi, Tokyo 184-8584, Japan}

%Shuhei Koyama
\author{Shuhei Koyama}
\affiliation{Institute of Astronomy, Graduate School of Science, The University of Tokyo, 2-21-1 Osawa, Mitaka, Tokyo 181-0015, Japan}

%Yusei Koyama
\author[0000-0002-0479-3699]{Yusei Koyama}
%\affiliation{Subaru Telescope, National Astronomical Observatory of Japan, National Institutes of Natural Sciences (NINS), 650 North Aohoku Place, Hilo, HI 96720, USA}
\affiliation{Subaru Telescope, National Astronomical Observatory of Japan, National Institutes of Natural Sciences (NINS), 650 North A'ohoku Place, Hilo, HI 96720, USA}
\affiliation{Department of Astronomical Science, SOKENDAI (The Graduate University for Advanced Studies), Osawa 2-21-1, Mitaka, Tokyo, 181-8588, Japan}

%Chien-Hsiu Lee
\author[0000-0003-1700-5740]{Chien-Hsiu Lee} 
\affiliation{W. M. Keck Observatory, Kamuela, HI 96743, USA}

%Akinori Matsumoto
\author{Akinori Matsumoto}
\affiliation{Institute for Cosmic Ray Research, The University of Tokyo, 5-1-5 Kashiwanoha, Kashiwa, Chiba 277-8582, Japan}
\affiliation{Department of Physics, Graduate School of Science, The University of Tokyo, 7-3-1 Hongo, Bunkyo, Tokyo 113-0033, Japan}

%Ken Mawatari
\author[0000-0003-4985-0201]{Ken Mawatari}
%\affiliation{National Astronomical Observatory of Japan, Osawa 2-21-1, Mitaka, Tokyo 181-8588, Japan}
\affiliation{National Astronomical Observatory of Japan, 2-21-1 Osawa, Mitaka, Tokyo 181-8588, Japan}

%Takashi Moriya
\author[0000-0003-1169-1954]{Takashi J. Moriya}
%\affiliation{National Astronomical Observatory of Japan, National Institutes of Natural Sciences, 2-21-1 Osawa, Mitaka, Tokyo 181-8588, Japan}
\affiliation{National Astronomical Observatory of Japan, 2-21-1 Osawa, Mitaka, Tokyo 181-8588, Japan}
\affiliation{School of Physics and Astronomy, Faculty of Science, Monash University, Clayton, Victoria 3800, Australia}

%Kentaro Motohara
\author{Kentaro Motohara}
%\affiliation{National Astronomical Observatory of Japan, National Institutes of Natural Sciences, 2-21-1 Osawa, Mitaka, Tokyo 181-8588, Japan}
\affiliation{National Astronomical Observatory of Japan, 2-21-1 Osawa, Mitaka, Tokyo 181-8588, Japan}
\affiliation{Institute of Astronomy, Graduate School of Science, The University of Tokyo, 2-21-1 Osawa, Mitaka, Tokyo 181-0015, Japan}

%Kai Murai
\author{Kai Murai}
\affiliation{Institute for Cosmic Ray Research, The University of Tokyo, 5-1-5 Kashiwanoha, Kashiwa, Chiba 277-8582, Japan}

%Moka Nishigaki
\author{Moka Nishigaki}
\affiliation{National Astronomical Observatory of Japan, 2-21-1 Osawa, Mitaka, Tokyo 181-8588, Japan}
\affiliation{Department of Astronomical Science, SOKENDAI (The Graduate University for Advanced Studies), Osawa 2-21-1, Mitaka, Tokyo, 181-8588, Japan}

%Masato Onodera
\author[0000-0003-3228-7264]{Masato Onodera}
\affiliation{Subaru Telescope, National Astronomical Observatory of Japan, National Institutes of Natural Sciences (NINS), 650 North A'ohoku Place, Hilo, HI 96720, USA}
\affiliation{Department of Astronomical Science, SOKENDAI (The Graduate University for Advanced Studies), Osawa 2-21-1, Mitaka, Tokyo, 181-8588, Japan}

%Yoshiaki Ono
\author[0000-0001-9011-7605]{Yoshiaki Ono}
\affiliation{Institute for Cosmic Ray Research, The University of Tokyo, 5-1-5 Kashiwanoha, Kashiwa, Chiba 277-8582, Japan}

%Michael Rauch
\author{Michael Rauch}
\affiliation{Carnegie Observatories, 813 Santa Barbara Street, Pasadena, CA 91101, USA}

%Tomoki Saito
\author{Tomoki Saito}
\affiliation{Nishi-Harima Astronomical Observatory, Centre for Astronomy, University of Hyogo, 407-2 Nishigaichi, Sayo, Sayo-gun, Hyogo 679-5313}

%Rin Sasaki
\author{Rin Sasaki}
\affiliation{Department of Physical Science and Materials Engineering, Faculty of Science and Engineering, Iwate University \\
3-18-34 Ueda, Morioka, Iwate 020-8550, Japan}

%Akihiro Suzuki
\author[0000-0002-7043-6112]{Akihiro Suzuki}
\affiliation{Research Center for the Early Universe, The University of Tokyo, 7-3-1 Hongo, Bunkyo, Tokyo 113-0033, Japan}

%Tsutomu Takeuchi
\author[0000-0001-8416-7673]{Tsutomu T.\ Takeuchi}
\affiliation{Division of Particle and Astrophysical Science, Nagoya University, Furo-cho, Chikusa-ku, Nagoya 464--8602, Japan}
\affiliation{The Research Center for Statistical Machine Learning, the Institute of Statistical Mathematics, 10-3 Midori-cho, Tachikawa, Tokyo 190---8562, Japan}

%Hiroya Umeda
\author{Hiroya Umeda}
\affiliation{Institute for Cosmic Ray Research, The University of Tokyo, 5-1-5 Kashiwanoha, Kashiwa, Chiba 277-8582, Japan}
\affiliation{Department of Physics, Graduate School of Science, The University of Tokyo, 7-3-1 Hongo, Bunkyo, Tokyo 113-0033, Japan}

%Masayuki Umemura
\author{Masayuki Umemura}
%\affiliation{Center for Computational Sciences, University of Tsukuba, Tsukuba, Ibaraki 305-8577, Japan}
\affiliation{Center for Computational Sciences, University of Tsukuba, Ten-nodai, 1-1-1 Tsukuba, Ibaraki 305-8577, Japan}

%Kuria Watanabe
\author{Kuria Watanabe}
\affiliation{Department of Astronomical Science, SOKENDAI (The Graduate University for Advanced Studies), Osawa 2-21-1, Mitaka, Tokyo, 181-8588, Japan}

%Kiyoto Yabe
\author[0000-0001-6229-4858]{Kiyoto Yabe}
\affiliation{Kavli Institute for the Physics and Mathematics of the Universe (WPI), University of Tokyo, Kashiwa, Chiba 277-8583, Japan}

%Yechi Zhang
\author{Yechi Zhang}
\affiliation{Institute for Cosmic Ray Research, The University of Tokyo, 5-1-5 Kashiwanoha, Kashiwa, Chiba 277-8582, Japan}
\affiliation{Department of Physics, Graduate School of Science, The University of Tokyo, 7-3-1 Hongo, Bunkyo, Tokyo 113-0033, Japan}

\begin{abstract}
We present demography of the dynamics and gas-mass fraction of 33 extremely metal-poor galaxies (EMPGs) with metallicities of $0.015-0.195~Z_\odot$ and low stellar masses of $10^4-10^8~\mathrm{M_\odot}$ in the local universe.
We conduct deep optical integral-field spectroscopy (IFS) for the low-mass EMPGs with the medium high resolution ($R=7500$) grism of the 8m-Subaru FOCAS IFU instrument by the EMPRESS 3D survey, and investigate H$\alpha$ emission of the EMPGs.
Exploiting the resolution high enough for the low-mass galaxies, we derive gas dynamics with the H$\alpha$ lines by the fitting of 3-dimensional disk models. We obtain an average maximum rotation velocity ($\vmax$) of $15\pm3~\mathrm{km~s^{-1}}$ and an average intrinsic velocity dispersion ($\sigma_0$) of $27\pm10~\mathrm{km~s^{-1}}$ for 15 spatially resolved EMPGs out of the 33 EMPGs, and find that all of the 15 EMPGs have $\voversigma<1$ suggesting dispersion dominated systems. 
There is a clear decreasing trend of $\voversigma$ with the decreasing stellar mass and metallicity. 
We derive the gas mass fraction ($f_\mathrm{gas}$) for all of the 33 EMPGs, 
and find no clear dependence on stellar mass and metallicity.
These $\voversigma$ and $\fgas$ trends should be compared with young high-$z$ galaxies observed by the forthcoming JWST IFS programs to understand the physical origins of the EMPGs in the local universe.

\end{abstract}

\keywords{galaxies: dwarf --- galaxies: evolution --- galaxies: kinematics and dynamics}

\section{Introduction}
Classical galaxy formation theory suggests galaxy initially forms as angular-momentum supported disks \citep[][]{White+78,Fall+80,Blumenthal+84}.
Primordial galaxies evolve into the various types of galaxies we see today involving complex interplay between different processes: accretion of cold gas, minor and major mergers, stellar and active-galactic-nuclei feedback.
One way to understand the evolution of primordial galaxies and the complex interplay is to study the dynamics of high-$z$ galaxies.
\cite{Rizzo+20,Rizzo+21} analyze the kinematics of $z\sim4-5$ galaxies and obtain large maximum rotation velocity ($\vmax$) that is $\sim$10 times the velocity dispersion ($\sigma_0$).
\cite{Tokuoka+22} identifies a $z\sim9$ galaxy that possibly presents clear rotation. 
\revise{Given that galaxies in \cite{Rizzo+20} and \cite{Tokuoka+22} reside in low-mass dark matter (DM) halos with halo masses of $\sim10^{10}~M_\mathrm{\odot}$ and $\sim10^9~M_\mathrm{\odot}$, respectively, the disk structure may not be stable but easily disrupted by inflow or outflow.}
Simulations of \cite{Dekel+20} suggest a critical \revise{DM} halo mass of $M_\mathrm{h}<2\times10^{11}~\mathrm{M_\odot}$, below which disk cannot survive and the galaxy becomes dispersion-dominated.

Although the dynamics of high-$z$ galaxies can be investigated with the state-of-the-art observation facilities (e.g., James Webb Space Telescope, hereafter JWST), local dwarf galaxies with recent starbursts are also important test-beds of galaxy evolution theories due to their apparent brightness.
Local galaxies are advantageous for conducting deep observations with high spectral and spatial resolutions, such as optical integral filed spectroscopy.
The spatially resolved H$\alpha$ recombination line can be used to trace the kinematics of ionized gas \citep[e.g.,][]{Green+14,Barat+20}.
Among local dwarf galaxies, EMPGs are considered as local counterparts of high-$z$ primordial galaxies having gas-phase metallicity (hereafter metallicity) below $10\%~\mathrm{Z_\odot}$.
EMPGs typically have low stellar masses $\lesssim10^8~\mathrm{M_\odot}$ and high specific star formation rates $\gtrsim1~\mathrm{Gyr}^{-1}$ suggestive of shallow gravitational potential and recent starburst, respectively.
Kinematics of EMPGs can provide us a hint of the important mechanism (e.g., inflow/outflow) during the early stage of galaxy formation.
\revise{Despite that EMPGs may differ from primordial galaxies on numerous aspects (star formation histories, stellar population, etc.), we aim to provide a clear correlation between dynamics and metallicity/stellar mass that can be extrapolated to high-$z$ primordial galaxies.}
Two important quantities indicating the detailed gas dynamical state are the relative level of rotation, via the $\voversigma$ ratio, and the mass composition, via the gas mass fraction ($\fgas$).

Recently, a project ``Extremely Metal-Poor Representatives Explored by the Subaru Survey (EMPRESS)'' has been launched (\citealt{Kojima+20}, here after Paper I). EMPRESS aims to select faint EMPG photometric candidates from Subaru/Hyper Suprime-Cam (HSC; \citealt{Miyazaki+18}) deep optical ($i_\mathrm{lim} = 26~\mathrm{mag}$; \citealt{Aihara+19}) images, which are 2 dex deeper than those of SDSS. Conducting follow-up spectroscopic observations of the EMPG photometric candidates, EMPRESS has identified new 12 EMPGs with low stellar masses of $10^{4.2}–10^{6.6}~\mathrm{M_\odot}$ (Papers I, IV; \citealt{Nakajima+22}, hereafter Paper V; \citealt{Xu+22}, hereafter Paper VI). Remarkably, J1631+4426 has been reported to have a metallicity of 0.016 $\mathrm{Z_\odot}$, which is the lowest metallicity identified so far (Paper I; c.f. \citealt{Thuan+22}).

This paper is the twelfth paper of EMPRESS, reporting a demography of H$\alpha$ kinematics of EMPGs observed with Subaru/Faint Object Camera and Spectrograph (FOCAS) IFU \citep[][]{Ozaki+20} in a series of the Subaru Intensive Program entitled EMPRESS 3D (PI: M. Ouchi). So far, EMPRESS has released 8 papers related to EMPGs, each of which reports the survey design (Paper I), high Fe/O ratios suggestive of massive stars (\citealt{Kojima+21}, hereafter Paper II; \citealt{Isobe+22a}, hereafter Paper IV), morphology (\citealt{Isobe+21}, hereafter Paper III), low-Z ends of metallicity diagnostics (Paper V), outflows (Paper VI), the shape of incident spectrum that reproduces high-ionization lines (\citealt{Umeda+22}, hereafter Paper VII), the primordial He abundance (\citealt{Matsumoto+22}, hereafter Paper VIII), and pioneer results of H$\alpha$ kinematics (\citealt{EMPRESSIX}, hereafter Paper IX).

The paper is structured as follows. Section \ref{data} explains our observations and dataset. Section \ref{results} describes how we derive rotation velocity, velocity dispersion and gas mass fraction. We discuss and summarize our findings in Sections \ref{discuss} and \ref{sum}, respectively. Throughout the paper we \revise{assume a solar metallicity of 12+log(O/H) = 8.69 \citep[][]{Asplund+21} and} adopt a cosmological model with $H_0=70~\mathrm{km~s^{-1}~Mpc^{-1}}$, $\Omega_\Lambda =0.7$, and $\Omega_\mathrm{m} = 0.3$.

\section{Observations and Data Reductions}
\label{data}

\subsection{Galaxy Sample}
We make a compilation of EMPGs whose metallicities are determined by the direct method in the EMPRESS project and similar studies that are \cite{Kniazev+03,Kniazev+04,Thuna&Izotov05,Izotov+01,Izotov+07,Izotov+09,Izotov+12,Izotov+12b,Izotov+18,Izotov+19,Izotov+20,Izotov+21,Morales-Luis+11,Skillman+13,Hirschauer+16,SA16,Hsyu+17,James+17,Senchyna+19}.
We select targets for our IFU observations from this compilation by their apparent brightness and visibility on the nights of observations.
We prioritize the targets with low metallicity or possibly complicated dynamical features (e.g., multiple clumps, broad emission lines).
Finally, we obtain 32 targets that have metallicity of $\logOH \sim 6.86-7.98$.
Only 3 out of the 32 targets have $\logOH \gtrsim 7.69$, while the others have $\logOH \lesssim 7.69$ that meets the criteria of EMPGs in the EMPRESS Project.

\subsection{Observations}
In the EMPRESS 3D project, we conducted observations for the 32 targets over 9 half-nights in 2021--2022.
We conducted the observations using FOCAS IFU mounted on Subaru Telescope.
We took science frames using the low-resolution ($R\sim900$) 300B grism and the mid-high-resolution ($R\sim7500$) VPH680 grism (hereafter low- and high-resolution data, respectively).
The low-resolution data were successfully taken for all the 32 targets (see Kimihiko Nakjima et al. in prep.).
Paper IX reports the first six targets with high-resolution data taken in 2021 that enables us to study the dynamics of EMPGs.
In 2022, we further obtained high-resolution data for 20 targets.
In total, we obtained low-resolution data for 32 objects and high-resolution data for 26 ($=6+20$) targets.

Here we describe the observations in 2022 when we observed 21 targets.
Because of the relatively high redshift of J1234+3901, we used the VPH850 grim ($R\sim1350$) to take the high-resolution data.
The observation nights were April 20th, 21st, 22nd, and October 17th, 2022, with typical seeing sizes of $0.6''$, $0.5''$, $0.8''$, and $0.4''$, respectively.
There were thin clouds \revise{at} the beginning of observations on April 21st.
On the other nights, the sky was clear.
We took calibration data for flat fielding and wavelength calibration in the beginning of observations.
We observed standard stars in the beginning or at the end of the observations.
We found no detection in the high-resolution data of J1044+6306 and successfully obtained high-resolution data for 20($=21-1$) targets.
The observations are summarized in Table \ref{tab:obs}.

\begin{deluxetable*}{lccccc}
    \tablecaption{Summary of medium-high resolution FOCAS-IFU observations}
    \tablewidth{0pt}
    \tablehead{
        % \colhead{} & \colhead{} & \colhead{} & \colhead{} & 
        % \multicolumn{2}{c}{Low-resolution grism} & \multicolumn{2}{c}{High-resolution grism} \\
        % \colhead{EW$_0(\mathrm{H}\beta)$} & \colhead{$12+\log(\mathrm{O/H})$} &
        \colhead{ID} & \colhead{R.A.} & \colhead{Decl.} & \colhead{Redshift} &
        % \colhead{Date of Observation} & \colhead{Exposure} & 
        \colhead{Date of Observation} & \colhead{Exposure} \\
        \colhead{} & \colhead{(hhmmss)} & \colhead{(ddmmss)} & \colhead{} &
        % \colhead{(UT)} & \colhead{(sec)} & 
        \colhead{(UT)} & \colhead{(sec)}
    }
    \startdata
J0036+0052 & 00:36:30.40 & $+$00:52:34.71 & 0.0282 & Oct 18, 2022 & 1200 \\
J0057-0941 & 00:57:57.32 & $-$09:41:19.20 & 0.0150 & Oct 18, 2022 & 1200 \\
J0125+0759 & 01:25:34.19 & $+$07:59:24.69 & 0.0098 & Oct 18, 2022 & 1200 \\
J0159+0751 & 01:59:52.75 & $+$07:51:48.80 & 0.0611 & Oct 18, 2022 & 1200 \\
J0228-0210 & 02:28:02.59 & $-$02:10:55.55 & 0.0414 & Oct 18, 2022 & 1200 \\
SBS 0335-052E$^\mathrm{\textcolor{blue}{\dag}}$ & 03:37:44.06 & $-$05:02:40.19 & 0.0135 & Nov 25, 2021 & 1200 \\
J0811+4730 & 08:11:52.12 & $+$47:30:26.24 & 0.0445 & Apr 21, 2022 & 1200 \\
HS 0822+3542$^\mathrm{\textcolor{blue}{\dag}}$ & 08:25:55.44 & $+$35:32:31.92 & 0.0023 & Dec 14, 2021 & 1200 \\
J0840+4707 & 08:40:29.90 & $+$47:07:10.30 & 0.0422 & Apr 23, 2022 & 1200 \\
I Zw 18$^\mathrm{\textcolor{blue}{\dag}}$ & 09:34:02.03 & $+$55:14:28.07 & 0.0024 & Dec 14, 2021 & 180 \\
J0935-0115 & 09:35:39.20 & $-$01:15:41.41 & 0.0160 & Apr 22, 2022 & 1200 \\
J0943+3326 & 09:43:32.43 & $+$33:26:58.00 & 0.0018 & Apr 21, 2022 & 1200 \\
DDO 68 & 09:56:46.05 & $+$28:49:43.78 & 0.0019 & Apr 22, 2022 & 1200 \\
J1016+3754 & 10:16:24.53 & $+$37:54:45.97 & 0.0040 & Apr 21, 2022 & 1200 \\
Leo P & 10:21:45.10 & $+$18:05:17.20 & 0.0010 & Apr 22, 2022 & 1200 \\
J1044+6306$^\mathrm{\textcolor{blue}{\ddag}}$ & 10:44:42.67 & $+$63:06:02.30 & 0.0033 & Apr 23, 2022 & 1200 \\
J1044+0353$^\mathrm{\textcolor{blue}{\dag}}$ & 10:44:57.79 & $+$03:53:13.15 & 0.0128 & Dec 14, 2021 & 1200 \\
J1234+3901 & 12:34:15.70 & $+$39:01:16.41 & 0.1333 & Apr 22, 2022 & 1200 \\
% J1253-0312 & 12:53:05.96 & $-$03:12:58.94 & 0.0227 & Jul 13, 2021$^\mathrm{\textcolor{blue}{\dag}}$ & 120 & -- & -- \\
% J1323-0132 & 13:23:47.46 & $-$01:32:51.94 & 0.0227 & Jul 13, 2021$^\mathrm{\textcolor{blue}{\dag}}$ & 600 & -- & -- \\
% J1355+4651 & 13:55:25.64 & $+$46:51:51.34 & 0.0284 & Aug 14, 2021$^\mathrm{\textcolor{blue}{\dag}}$ & 1200 & -- & -- \\
J1418+2102 & 14:18:51.12 & $+$21:02:39.74 & 0.0086 & Apr 22, 2022 & 1200 \\
J1423+2257 & 14:23:42.88 & $+$22:57:28.80 & 0.0328 & Apr 22, 2022 & 1200 \\
J1452+0241 & 14:52:55.28 & $+$02:41:01.31 & 0.0054 & Apr 23, 2022 & 1200 \\
J1631+4426$^\mathrm{\textcolor{blue}{\dag}}$ & 16:31:14.24 & $+$44:26:04.43 & 0.0313 & Aug 14, 2021 & 1200 \\
J1702+2120 & 17:02:39.88 & $+$21:20:08.91 & 0.0249 & Apr 23, 2022 & 1200 \\
J2104-0035 & 21:04:55.30 & $-$00:35:22.00 & 0.0047 & Oct 18, 2022 & 1200 \\
J2115-1734$^\mathrm{\textcolor{blue}{\dag}}$ & 21:15:58.33 & $-$17:34:45.09 & 0.0230 & Aug 14, 2021 & 1200 \\
J2136+0414 & 21:36:58.81 & $+$04:14:04.31 & 0.0169 & Oct 18, 2022 & 1200 \\
J2302+0049 & 23:02:10.00 & $+$00:49:38.78 & 0.0332 & Oct 18, 2022 & 1200
% J2314+0154 & 23:14:37.55 & $+$01:54:14.27 & 0.0327 & Jun 18, 2021$^\mathrm{\textcolor{blue}{\dag}}$ & 868 & -- & --
    \enddata
    % \tablenotetext{\textcolor{blue}{\dag}}{Kimihiko Nakajima et al. in prep.}
    \tablenotetext{\textcolor{blue}{\dag}}{Paper IX}
    \tablenotetext{\textcolor{blue}{\ddag}}{no detection.}
    \label{tab:obs}
\end{deluxetable*}

\begin{figure*}[t!]
    \centering
    \plotone{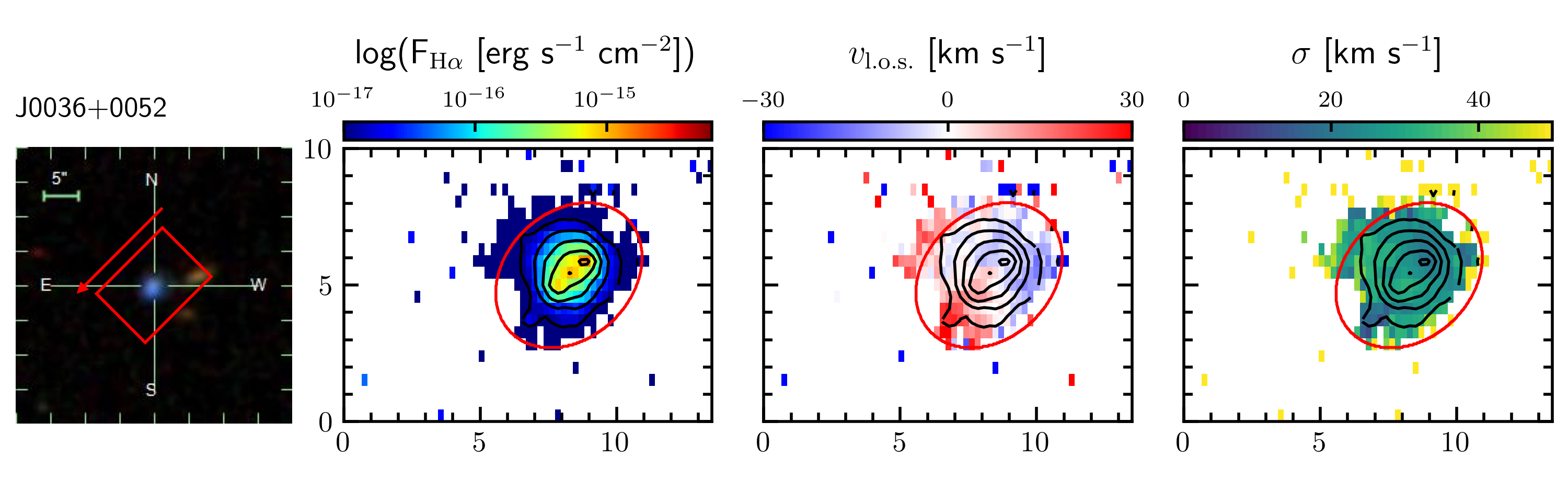}
    \plotone{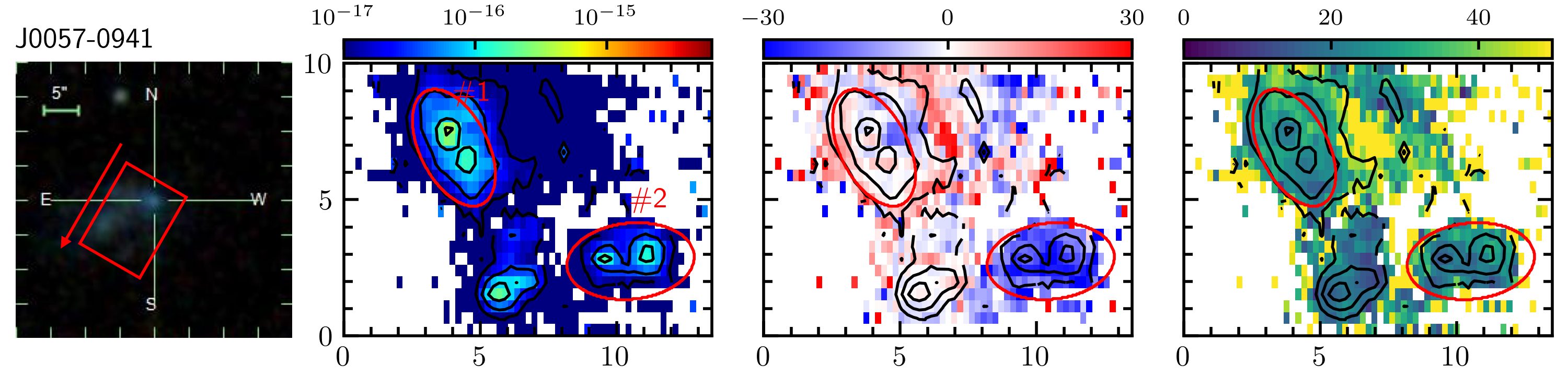}
    \plotone{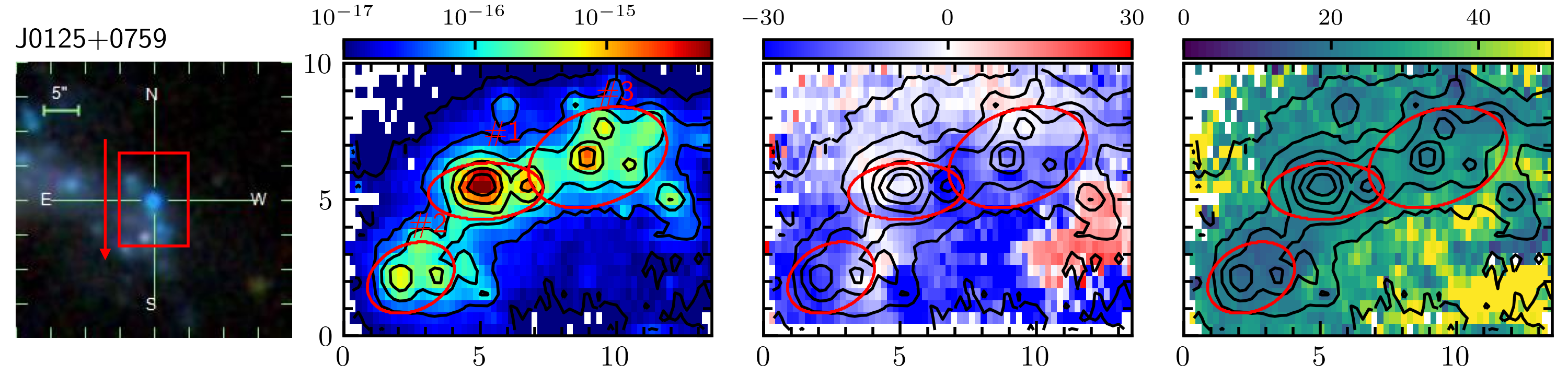}
    \plotone{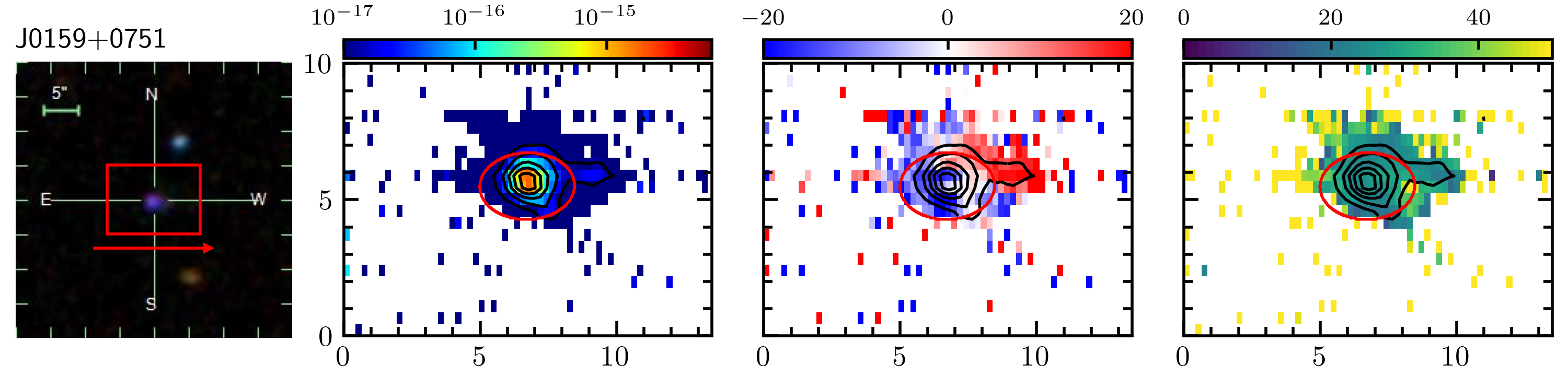}
    \plotone{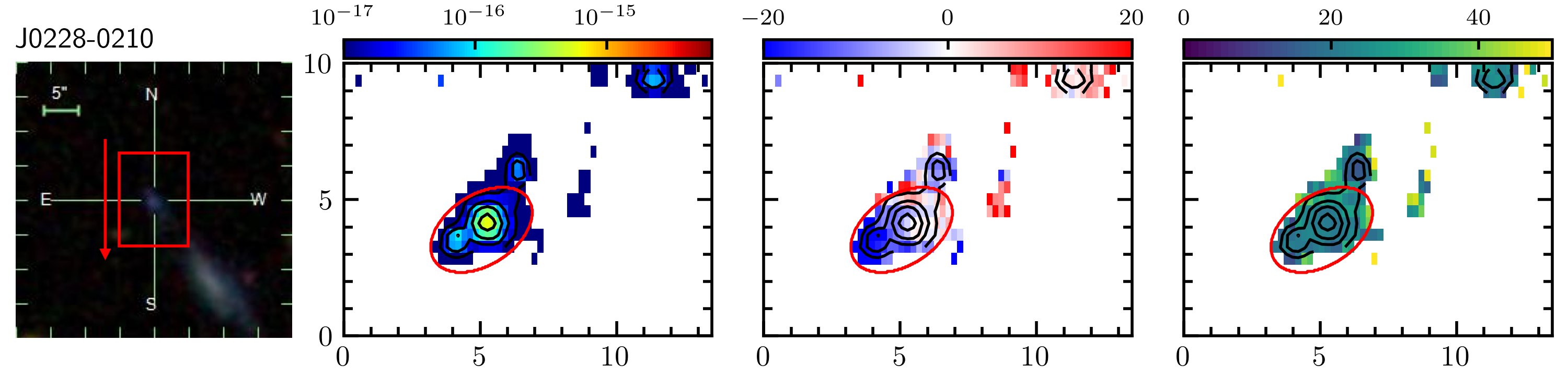}
    \caption{From left to right: SDSS cut-out, H$\alpha$ flux map, line-of-sight velocity map, and velocity dispersion map for each EMPG whose high-resolution data is reported in this study and Paper IX. Spaxels with S/N(H$\alpha$) $>3$ are plotted. The red rectangle on the SDSS cut-out indicates the pointing position of our IFU observations while the arrow indicates the direction of x-axis. The x- and y-axes are presented in arcsec. The black contours represent the H$\alpha$ flux in the range of $\log(F_\mathrm{H\alpha}/\mathrm{erg~s^{-1}~cm^{-2}})=-17$ to $-14.5$ with a step of 0.5 dex. The red circles highlight the apertures we obtained by source detection (see Section \ref{multi_clump}). Within the apertures we fit disk rotation models.}
    \label{fig:maps}
\end{figure*}

\subsection{Data reductions}

We use a reduction pipeline software of FOCAS IFU \citep{Ozaki+20} based on \texttt{PyRAF} (Tody 1986) and \texttt{Astropy} \citep{astropy:2013, astropy:2018, astropy:2022}.
The software performs bias subtraction, flat-fielding, wavelength calibration, cosmic ray removal, and flux calibration.
The software outputs three-dimensional (3D) data cubes of integral field spectroscopy (IFS) with and without sky background subtraction.
The 3D data cubes cover a filed-of-view (FoV) of $13.5''\times10''$ with $64\times23$ spaxels, which corresponds to a pixel scale of $0.215''/\mathrm{pix}$ and $0.435''/\mathrm{pix}$ in the x- and y-axis, respectively.
For the low-resolution data, the 3D data cubes cover the wavelengths of $3500 - 8000~\mathrm{\AA}$.
For the high-resolution data, the 3D data cubes cover the wavelengths of $6500 - 7500~\mathrm{\AA}$ and $6000 - 10000~\mathrm{\AA}$ for the VPH650 and VPH850 grim, respectively.
We estimate the flux uncertainties containing read-out noises and photon noises of sky and object emissions.

\subsection{Deblending of spatial components}
\label{multi_clump}

Our data includes EMPGs that have multiple spatial components as shown in the photometric images.
The multiple components can also be seen in the H$\alpha$ flux maps derived from our IFU data (see Figure \ref{fig:maps} and Section \ref{kinematics}).
In this study, we aim to discuss the properties of individual components exploiting the spatial resolution of the IFU data.
We define the components based on the morphology of the H$\alpha$ flux map using the tool of source detection and deblending in \texttt{Photutils}, an \texttt{Astropy} package.
We carefully choose the flux threshold to include only components that have photometric counterparts in the SDSS catalog.
Twenty targets appear to have only one component of our interest.
From 5 EMPGs which are J0057-0941, J2104-0035, J2115-1734, \revise{DDO 68, and I Zw 18}, we extract 2 components.
For J0125+0759, we extract 3 components.
We label the components with suffices, where the brightest one is labeled with \#1.
For \revise{I Zw 18}, we separate \revise{I Zw 18-NW} and \revise{I Zw 18}-SE.
For \revise{DDO 68}, we use the labels of \#2 and \#3 in consistency with the notations in previous studies \citep[e.g.,][]{Pustilnik+05}.
Although the origin of the multiple components in one system is debatable, we treat each component as an individual EMPG in this study, which means our sample consists of 33 ($=20+5\times2+3$) EMPGs in total.

\begin{figure*}[ht!]
    \centering
    \plotone{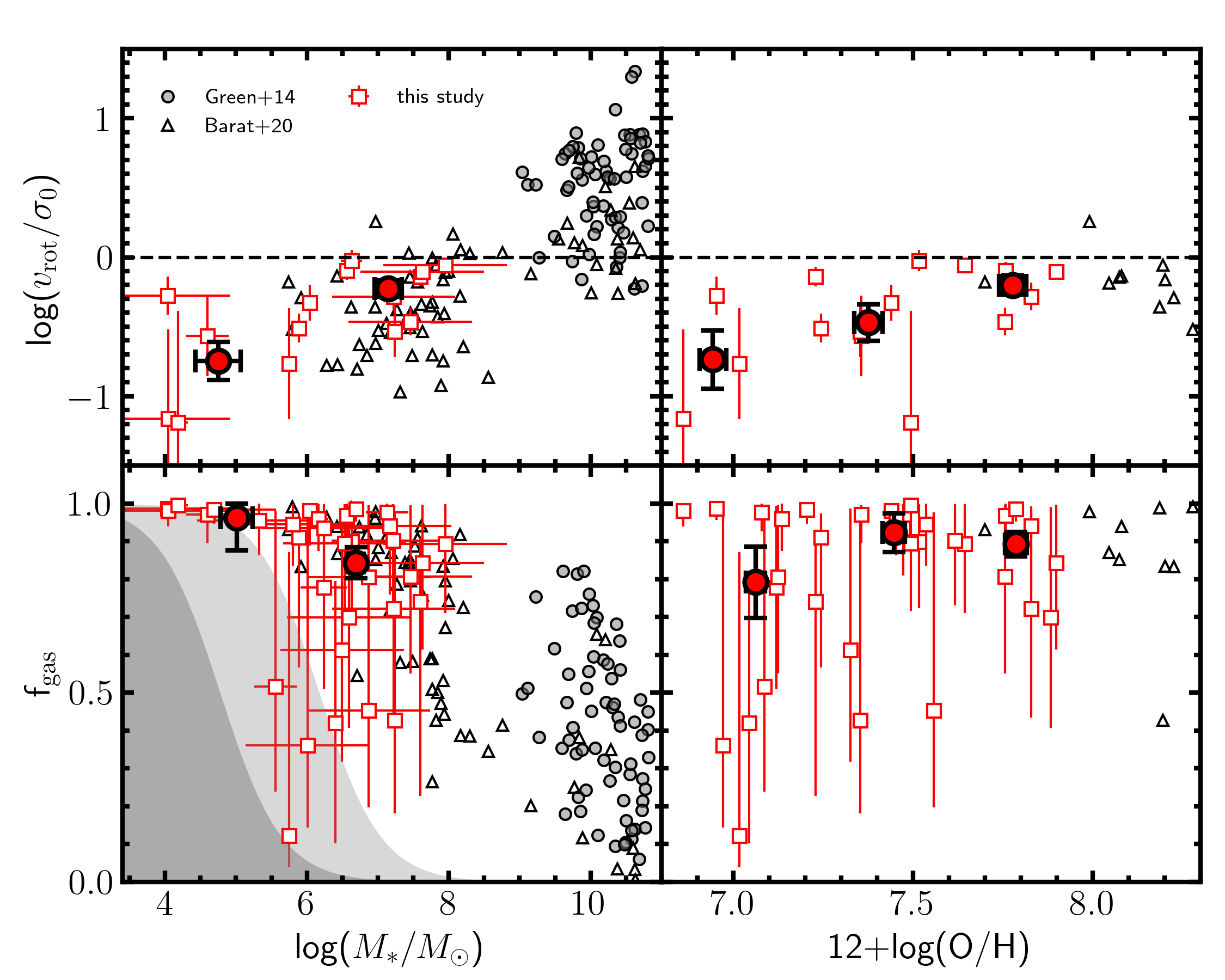}
    \caption{(Top) $v_\mathrm{max}/\sigma_0$ as a function of stellar mass (left) and gas-phase metallicity (right). The red squares are individual EMPGs while the red circles are average values in stellar mass (metallicity) bins. The grey circles and triangles are taken from the DYNAMO survey \citep[][]{Green+14} and the SH$\alpha$DE survey \citep[][]{Barat+20}, respectively. (Bottom) Same as the top panels but for $\fgas$. We plot the median values with red circles. The grey regions in the bottom left panel indicate the observational limit for $\fgas$ (see Section \ref{fgas}).}
    \label{fig:relations}
\end{figure*}

\section{Analyses and Results}
\label{results}

\subsection{Gas mass fraction}
\label{fgas}

Gas mass surface density ($\Sigma_\mathrm{gas}$) can be estimated from the SFR surface density adopting the Kennicutt-Schmidt law \citep[][]{Kennicutt98}:
\begin{align}
\Sigma_\mathrm{SFR}=&(2.5\pm0.7)\times10^{-4}\notag\\
&\left(\frac{\Sigma_\mathrm{gas}}{1~M_\odot~\mathrm{pc^{-1}}}\right)^{1.4\pm0.15}~M_\odot~\mathrm{yr^{-1}~kpc^{-2}}.
\label{eq:KS}
\end{align}
We estimate the SFR surface density ($\Sigma_\mathrm{SFR}$) from the H$\alpha$ flux using the relation from \cite{Kennicutt98}, assuming the initial mass function (IMF) of \cite{Chabrier03}:
\begin{equation}
\mathrm{SFR}~[M_\odot~\mathrm{yr^{-1}}]=4.4\times10^{-42}~L(\mathrm{H\alpha})~[\mathrm{erg~s^{-1}}].
\label{eq:SFR}
\end{equation}
We fit a Gaussian profile to the spectrum of emission line using the low-resolution data which offers a better signal-to-noise ratio (S/N) for the H$\alpha$ lines.

Gas mass fraction is defined as the ratio between the gas mass and the total baryonic mass within one effective radius:
\begin{equation}
f_\mathrm{gas}=\frac{M_\mathrm{gas}(\withinre)}{M_\mathrm{gas}(\withinre)+M_*(\withinre)}.
\label{eq:fgas_def}
\end{equation}
We sum up $\Sigma_\mathrm{gas}$ for the spaxles within $r_\mathrm{e}$ to obtain $M_\mathrm{gas}(\withinre)$.
The value of $r_\mathrm{e}$ is derived by \revise{fitting a Sersic profile to the H$\alpha$ flux maps.
The fitting procedure is integrated in the rotation disk models for the 15 spatially-resolved EMPGs (see Section \ref{kinematics}) and conducted separately using \texttt{galfit} \citep[][]{Peng+02,Peng+10} for the other 18($=33-15$) EMPGs.} 
For the stellar masses, there are 15 EMPGs whose stellar masses are measured in previous studies (Table \ref{tab:res}).
To estimate the stellar masses for the other EMPGs, we first obtain the $i-$band magnitudes for all the 33 EMPGs from the SDSS catalog.
We obtain an average mass-to-light ratio of 0.12 between the stellar mass and $i-$band magnitude using the 15 EMPGs with known stellar masses.
For the other 18($=33-15$) EMPGs, we derive the stellar masses from the absolute $i-$band magnitudes assuming the mass-to-light ratio.
We assume the effective radius of stellar component ($r_\mathrm{e,*}$) can be approximated by $r_\mathrm{e}$ and estimate $M_*(r<r_\mathrm{e})$ by dividing the stellar masses by 2.
\revise{We check the possible systematic errors given by our analysis method. We confirm that the gas mass fraction is consistent between the 15 galaxies whose stellar masses are derived from a fixed mass-to-light ratio and the rest of the sample. Since some EMPGs are identified from multiple components in one field of view (Section \ref{multi_clump}), we make sure the consistency between the $i-$band aperture and the spaxels used for $M_\mathrm{gas}(\withinre)$. Interestingly, we find the multiple components in one system generally have similar $\fgas$, implying that the multiple components may have similar star formation history.}

We evaluate the uncertainty of $M_*(\withinre)$ assuming $r_\mathrm{e,*}$ ranges from $r_\mathrm{e}/2$ to $2r_\mathrm{e}$.
We take the 0.3 dex scatter of Kennicutt-Schmidt law as the uncertainty of $M_\mathrm{gas}(\withinre)$.
The calibration from \cite{Kennicutt98} may underestimate the gas mass for metal-poor galaxies \citep[e.g.,][]{Shi+14}.
Therefore, we add an upper error of 1 dex to our estimation of $M_\mathrm{gas}(\withinre)$ following Paper IX.
We propagate these uncertainties to obtain the error of $\fgas$.
The results are summarized in Table \ref{tab:res}.
In the bottom panels of Figure \ref{fig:relations}, we show $f_\mathrm{gas}$ as a function of stellar mass and metallicity.
We obtain an median value of $\fgas\sim0.9$ larger than those of more massive galaxies in \cite{Barat+20}, which suggests EMPGs are likely to be gas-rich systems.

\revise{The average and standard deviation of $\fgas$ is 0.8 and 0.2, respectively. Six galaxies (J0057-0941, J0943+3326, J2104-0035, I Zw 18-NW/-SE, and Leo P) are 1$\sigma$ below the average having $\fgas<0.6$. One possibility of finding low $\fgas$ is the existence of old stellar population \citep[e.g., I Zw 18;][]{Vaduvescu+05,Aloisi+07}. We find Leo P may be relatively gas-deficient with $\fgas=0.12_{-0.08}^{+0.75}$ although with a large uncertainty.
\cite{Bernstein-Cooper+14} conduct H{\sc i} observations and obtain $\fgas\sim0.7$ for Leo P.
They claim the $\fgas$ value is relatively small among EMPGs, which is consistent with our results.}

%  TODO: discuss the galaxies with seemingly low fgas. TODO: also check ~5 galaxies with stellar mass from fixed mass-to-light ratio. TODO: discuss Leo P

For our EMPGs, we find no clear correlation between $\fgas$ and stellar mass or metallicity.
It may seem that only high $\fgas\gtrsim0.9$ can be found for EMPGs with $M_*<10^5~\mathrm{M_\odot}$.
However, the fact that apparently bright EMPGs are preferred by our sample selection may lead to selection bias towards high $\fgas$ EMPGs.
We consider the selection bias by simply assuming an H$\alpha$ surface brightness of $1\times10^{-17}~\mathrm{erg~s^{-1}~cm^{-2}~arcsec^{-2}}$, which is roughly the detection limit of our low-resolution data \citep[][]{Ozaki+20}.
We apply the H$\alpha$ surface brightness to Equations \ref{eq:KS} and \ref{eq:SFR} to derive the limiting $\Sigma_\mathrm{gas}$ for observable EMPGs.
\cite{Isobe+21} measure the effective radius of 27 EMPGs and obtain $r_\mathrm{e}\sim200^{+450}_{-110}~\mathrm{pc}$.
We take the minimum (maximum) value of $100~\mathrm{pc}$ ($1000~\mathrm{pc}$) to integrate the $\Sigma_\mathrm{gas}$ as the limiting gas mass.
Finally, we calculate the limiting $\fgas$ as a function of $M_*$ using Equation \ref{eq:fgas_def}.
In the bottom-left panel of Figure \ref{fig:relations}, we show two grey regions bellow the limit $\fgas$, where the lighter (darker) color corresponds to the case of minimum (maximum) $r_\mathrm{e}$ values.
We find the correlation between $\fgas$ and $M_*$ is possibly biased below $10^6~\mathrm{M_\odot}$.
On the other hand, there is no clear correlation between $\fgas$ and $M_*$ above $10^6~\mathrm{M_\odot}$.

\subsection{Gas kinematics}
\label{kinematics}

\begin{figure*}[t!]
    \centering
    \plotone{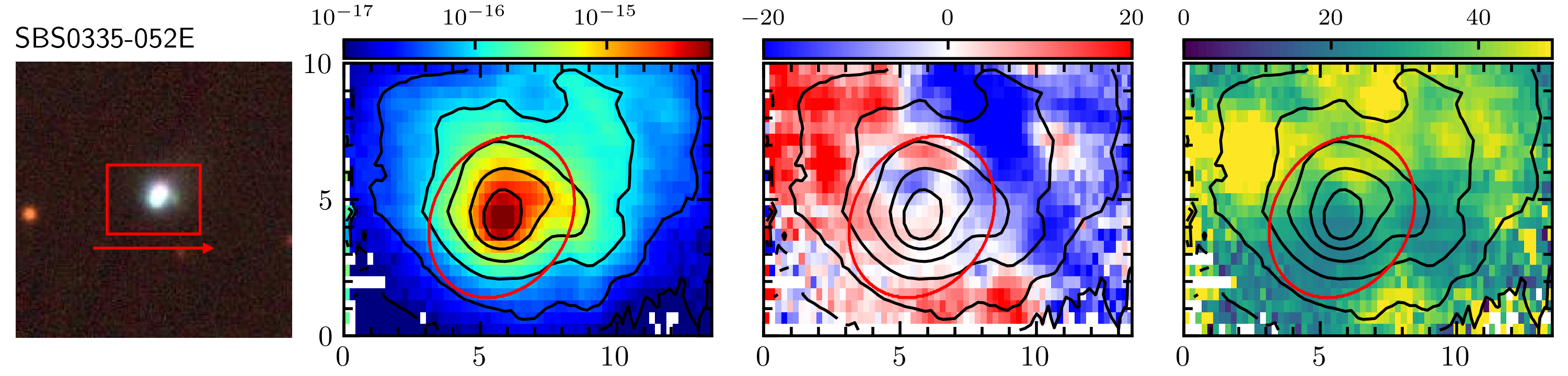}
    \plotone{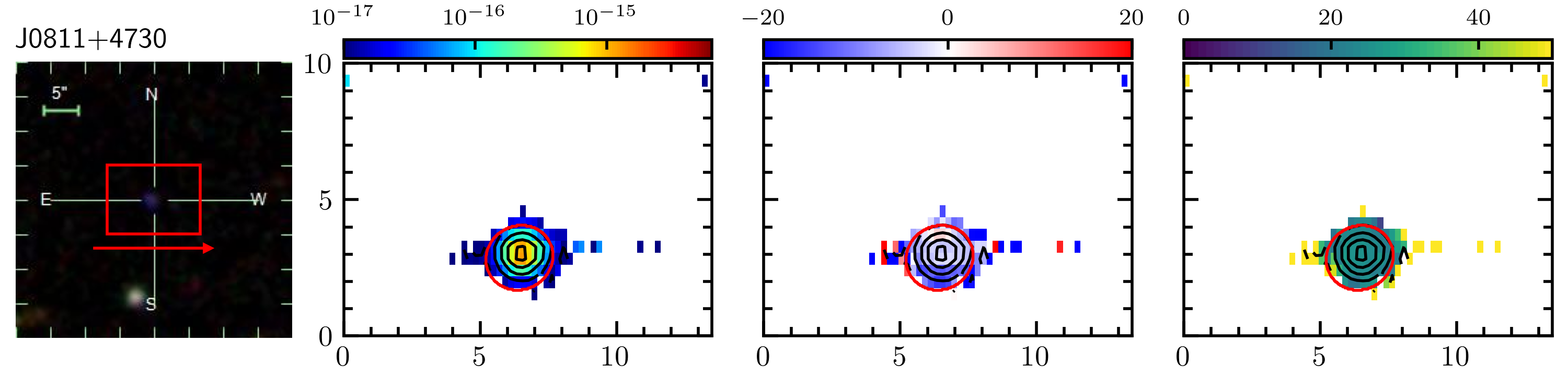}
    \plotone{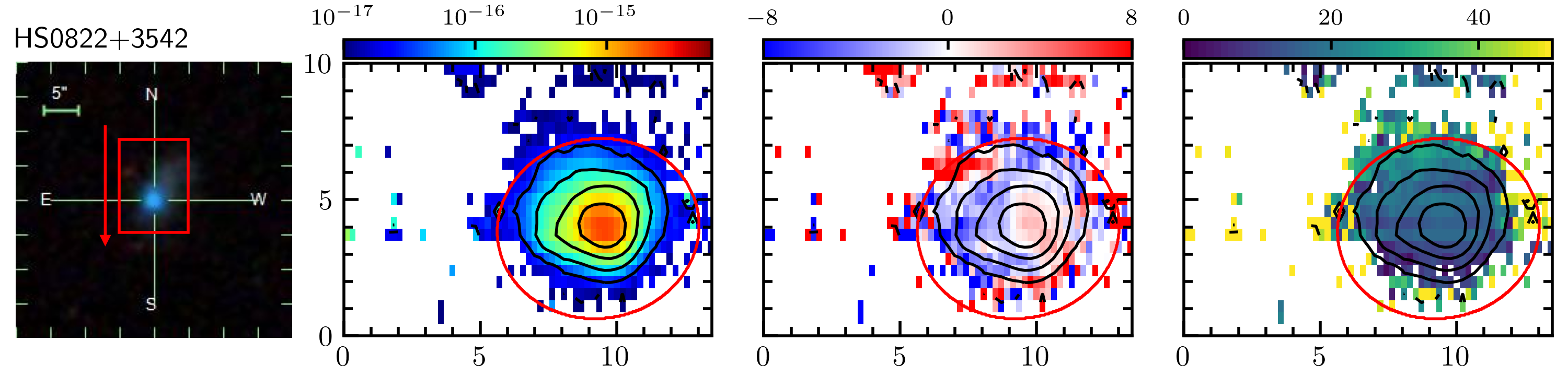}
    \plotone{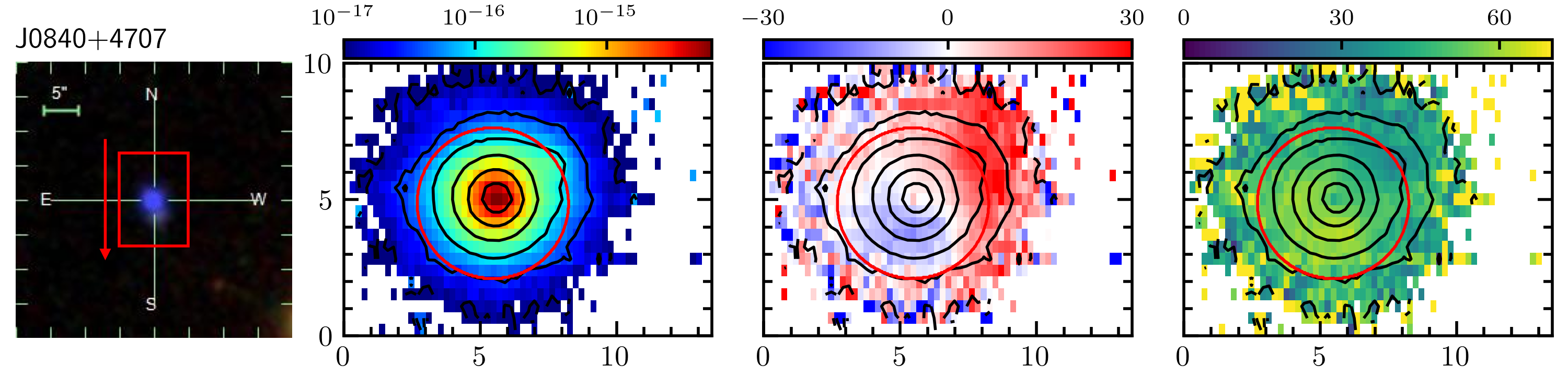}
    \plotone{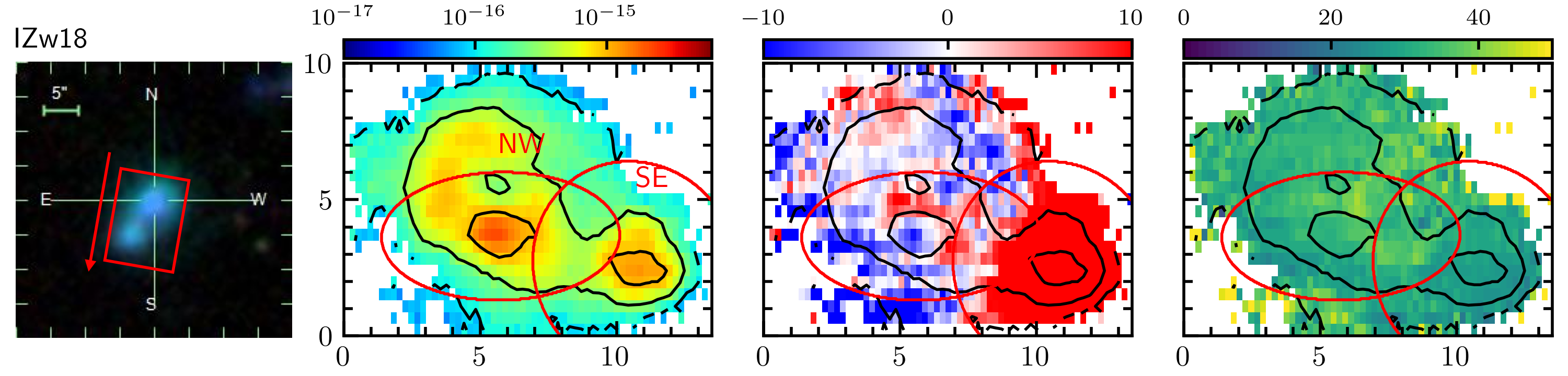}
    \plotone{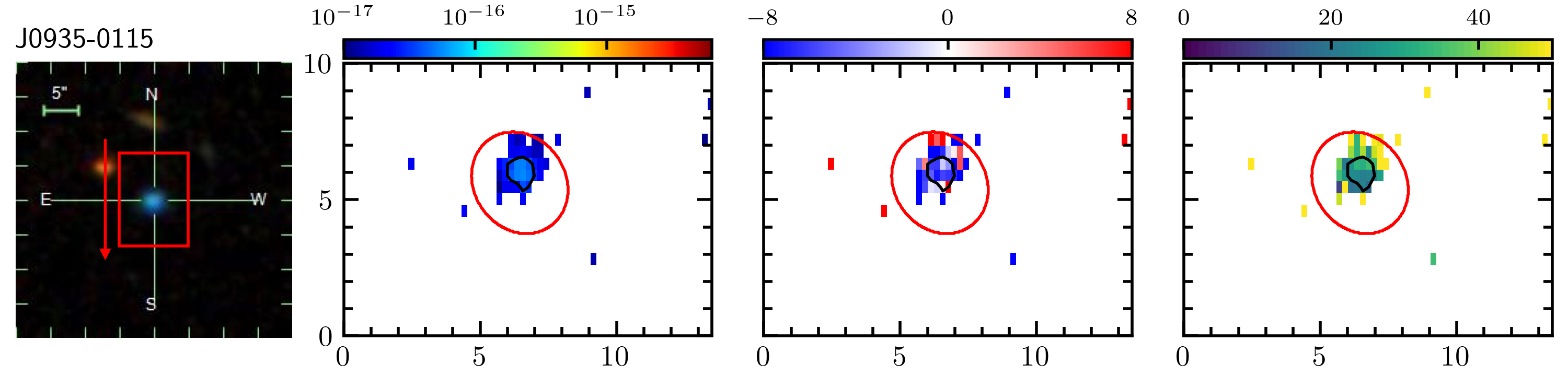}
    \caption{Figure \ref{fig:maps} continued.}
    \label{fig:maps2}
    \vspace{1cm}
\end{figure*}

\begin{figure*}[t!]
    \centering
    \plotone{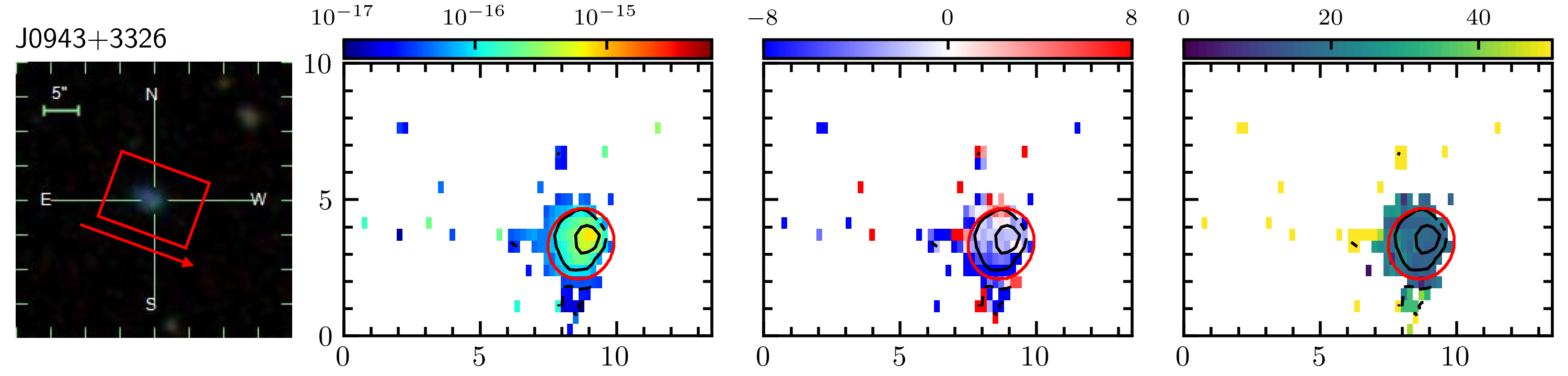}
    \plotone{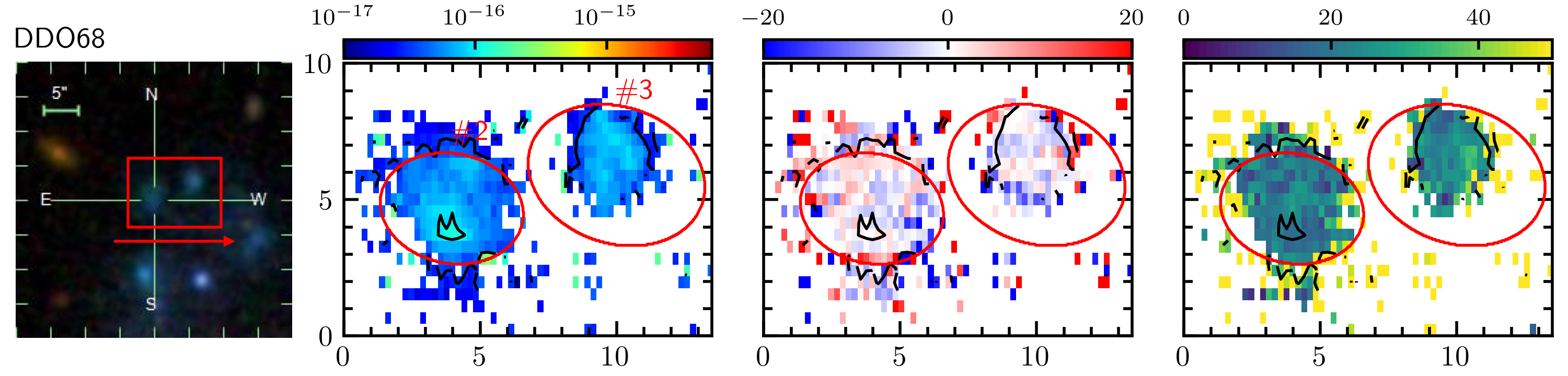}
    \plotone{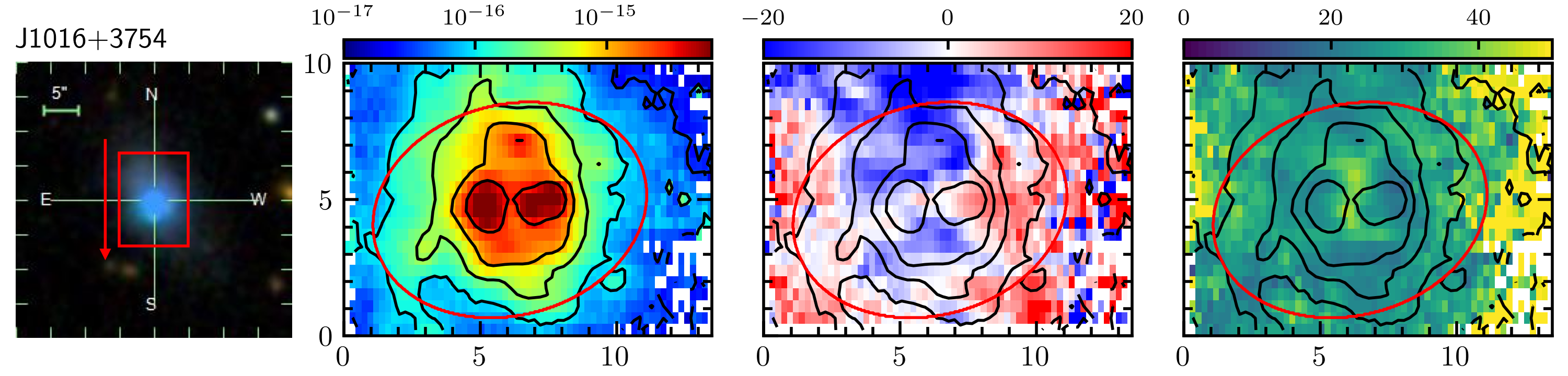}
    \plotone{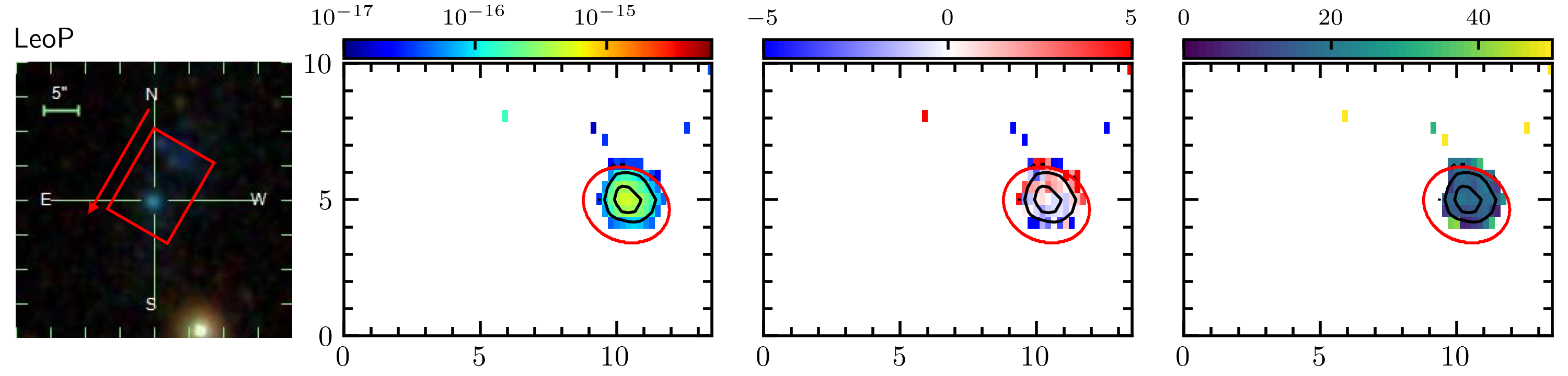}
    \plotone{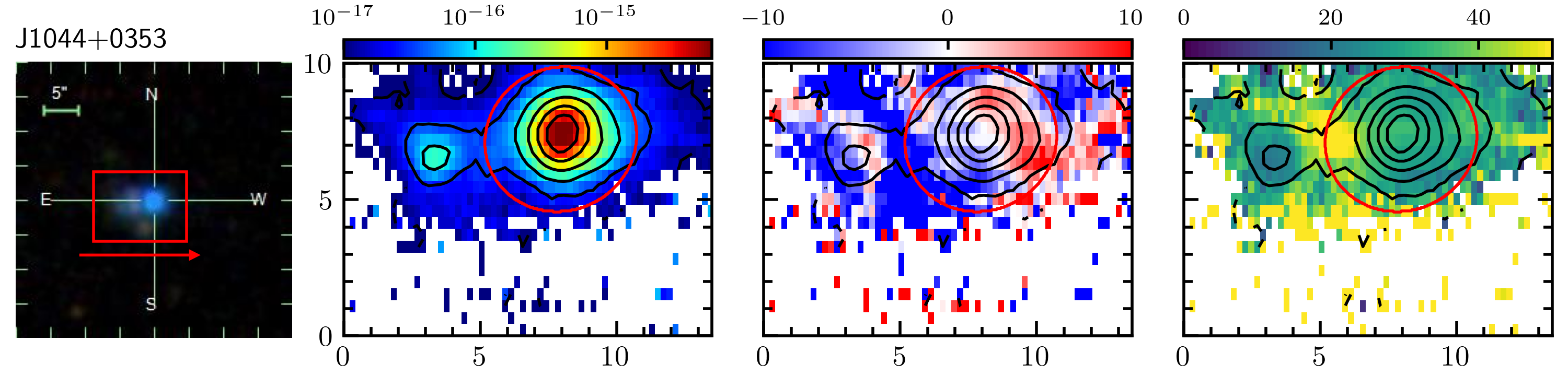}
    \plotone{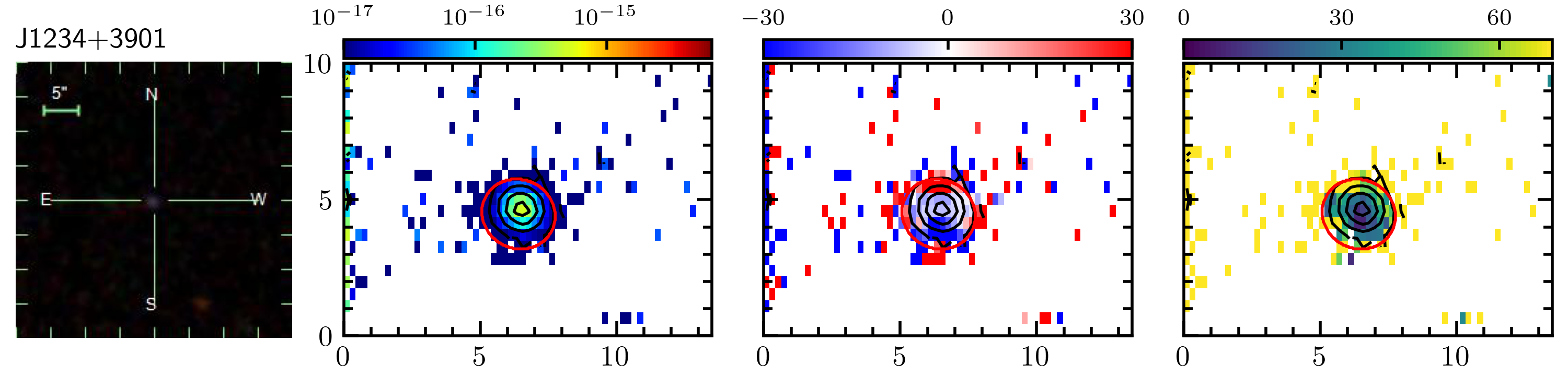}
    \caption{Figure \ref{fig:maps} continued.}
    \label{fig:maps3}
    \vspace{1cm}
\end{figure*}

\begin{figure*}[t!]
    \centering
    \plotone{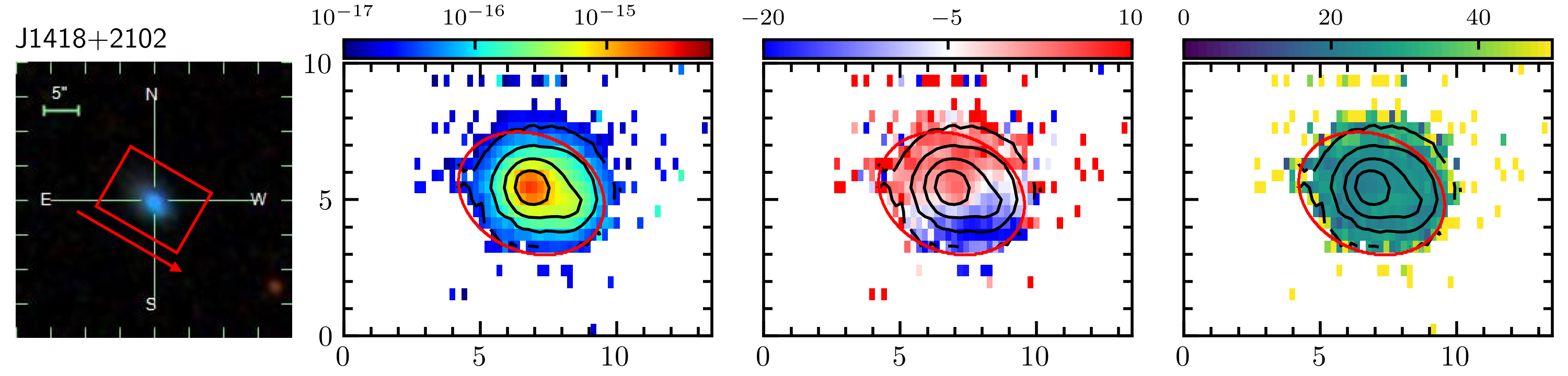}
    \plotone{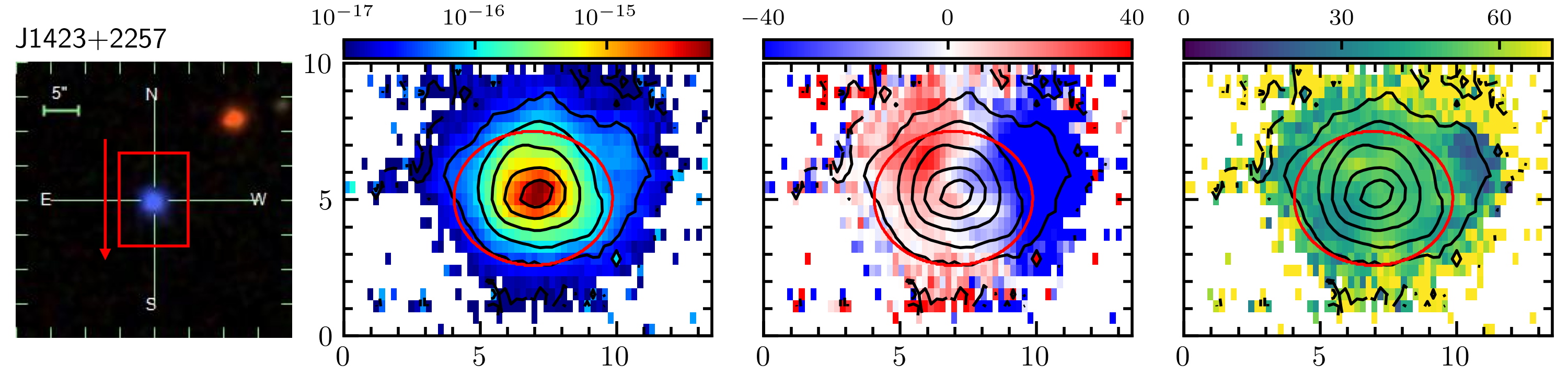}
    \plotone{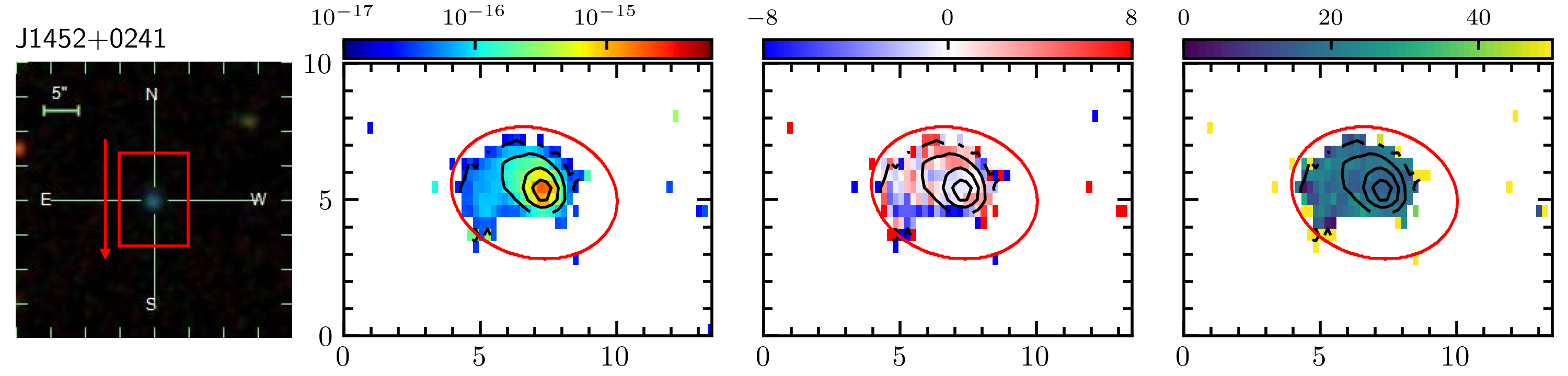}
    \plotone{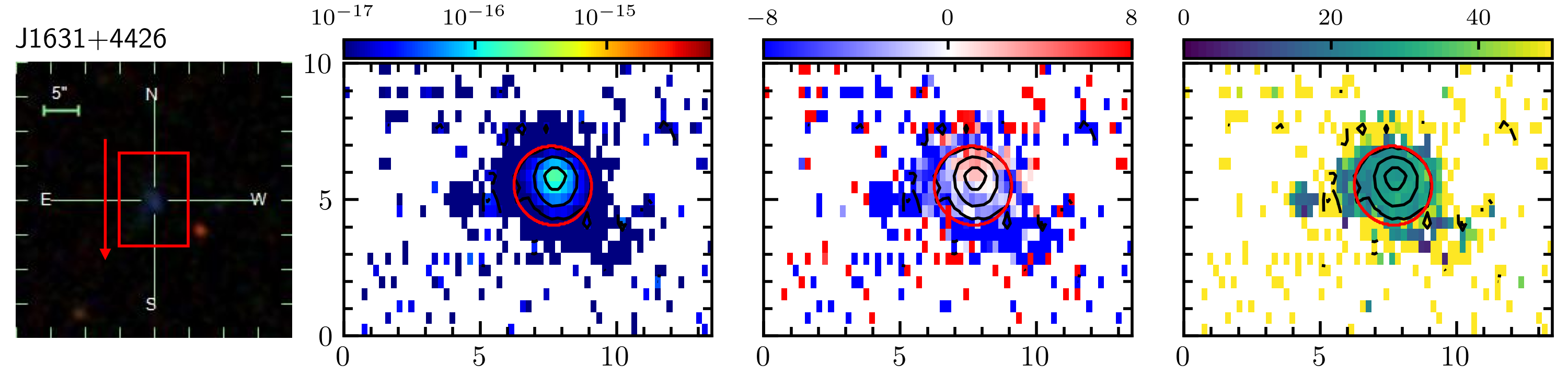}
    \plotone{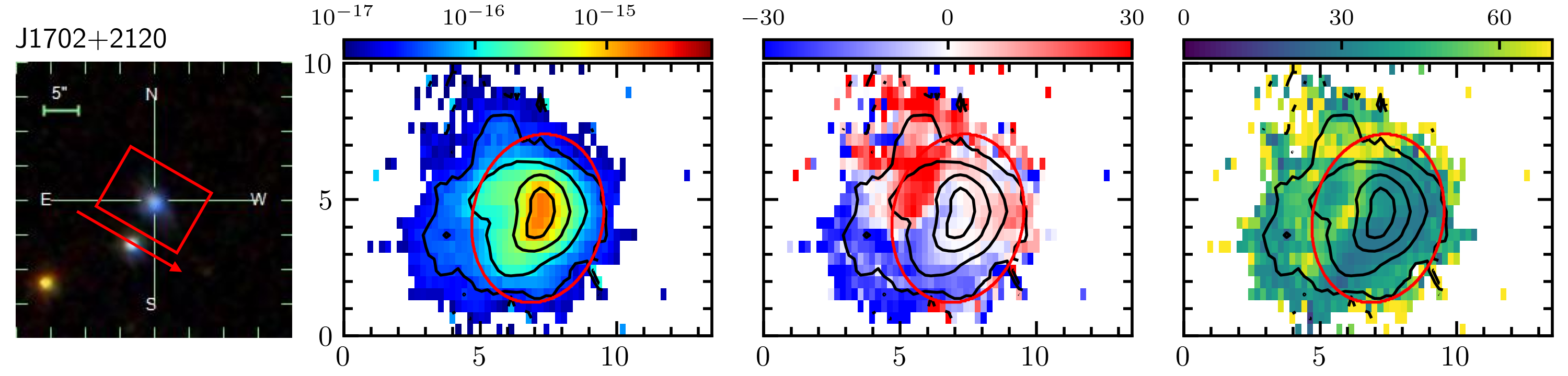}
    \plotone{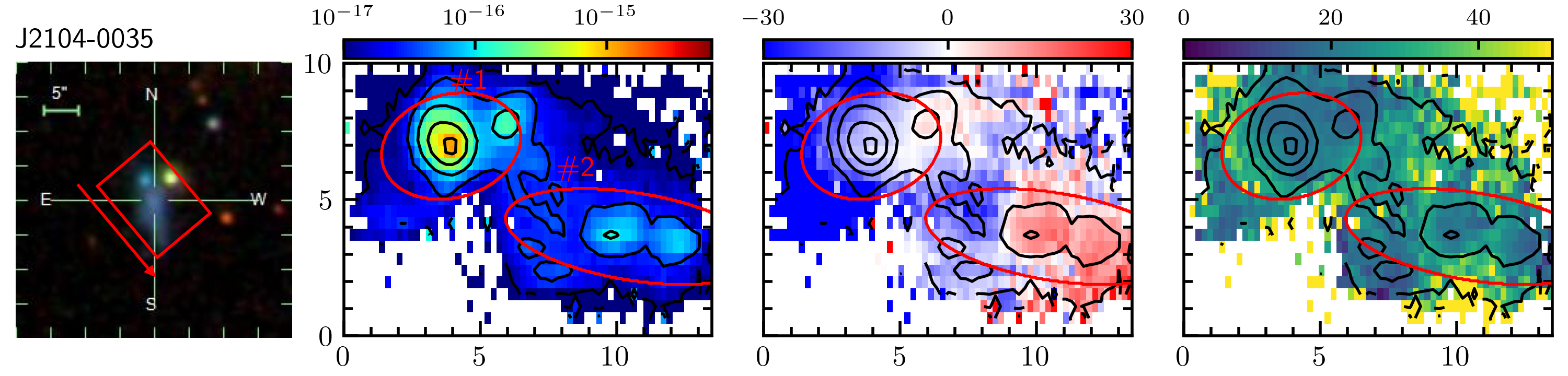}
    \caption{Figure \ref{fig:maps} continued.}
    \label{fig:maps4}
\end{figure*}

\begin{figure*}[t!]
    \centering    
    \plotone{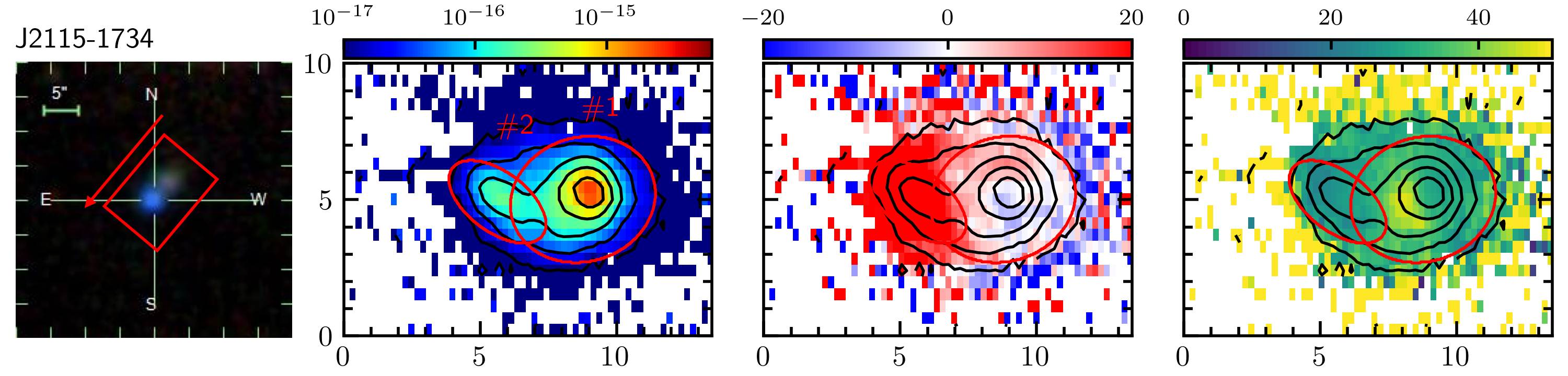}
    \plotone{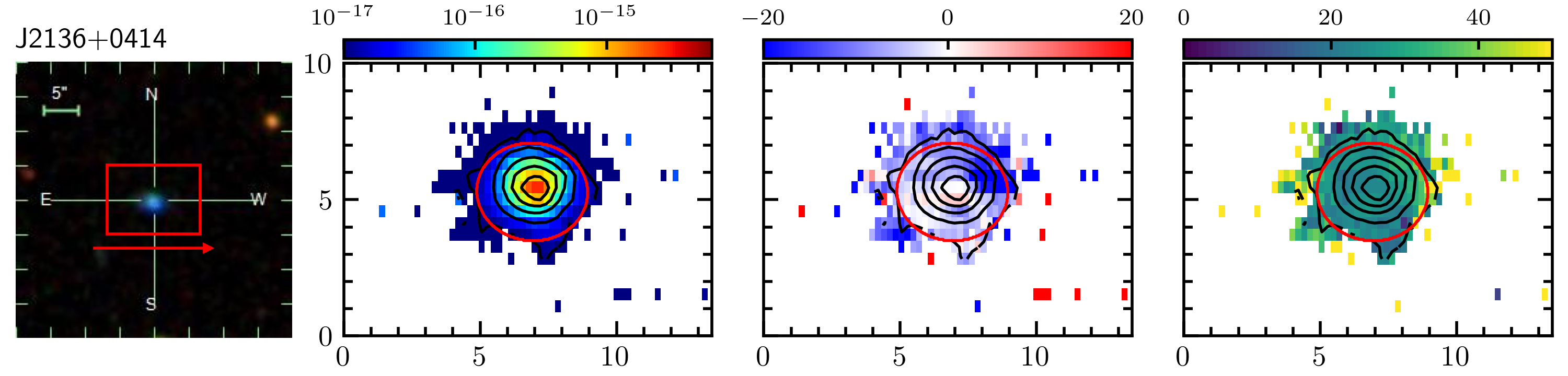}
    \plotone{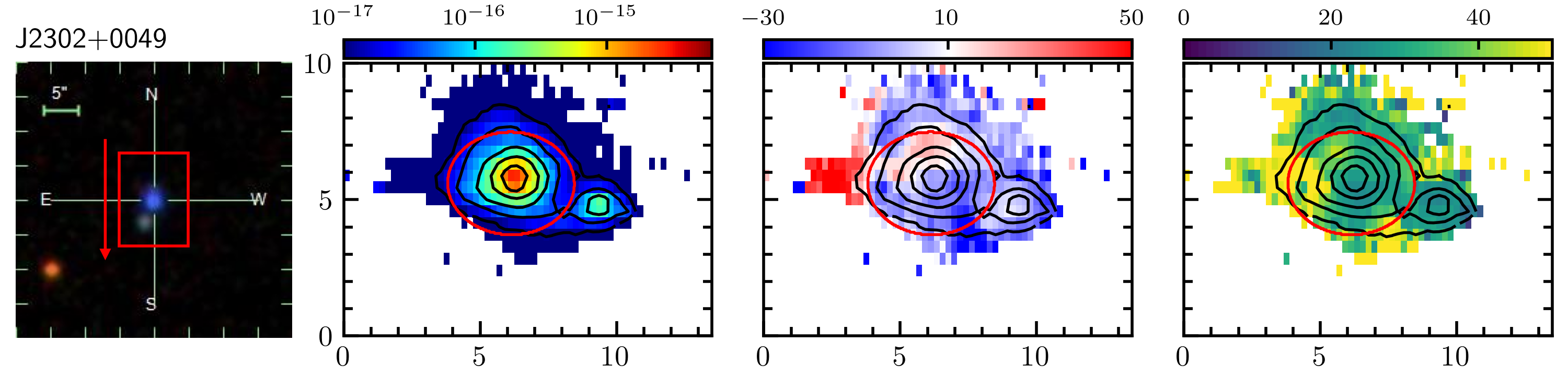}
    \caption{Figure \ref{fig:maps} continued.}
    \label{fig:maps5}
\end{figure*}

We derive the kinematical properties of our EMPGs from the H$\alpha$ lines.
The spatial distribution of line-of-sight velocity ($v_\mathrm{l.o.s.}$) and velocity dispersion ($\sigma$) can be derived by fitting a Gaussian profile to each spaxel of the high-resolution data.
For the line-of-sight velocity, we derive the velocity from the central wavelength ($\lambda_\mathrm{H\alpha}$):
\begin{equation}
    v_\mathrm{l.o.s.} = c(\lambda_\mathrm{H\alpha} - \lambda_0 - \lambda_\mathrm{shift})/\lambda_0,
\end{equation}
where $\lambda_0$ represents the systemic wavelength.
For EMPGs that can be fitted by a rotation disk model (see below), $\lambda_0$ is the observed wavelength at the center of the disk model.
Otherwise, we adopt the central wavelength in the spaxel where the flux of H$\alpha$ is largest among all the spaxels.
The value of $\lambda_\mathrm{shift}$ is given by the slit-width effect that is caused by a flux gradient parallel to the wavelength direction (see Section 3.2 in Paper IX) and $c$ is the speed of light.
For the velocity dispersion, we subtract the instrumental broadening ($\sigma_\mathrm{inst}$) % and thermal broadening ($\sigma_\mathrm{th}$) 
from the line width ($\sigma_\mathrm{H\alpha}$):
\begin{equation}
    \sigma = \sqrt{\sigma_\mathrm{H\alpha}^2 - \sigma_\mathrm{inst}^2}. 
\end{equation}
We estimate a typical value of $\sigma_\mathrm{inst}=17\pm2~\mathrm{km~s^{-1}}$ from the line widths of unresolved skylines, where the uncertainty is the standard deviation of $\sigma_\mathrm{inst}$ from different spaxels.
We show the H$\alpha$ flux, velocity and dispersion maps in Figures \ref{fig:maps} and \ref{fig:maps2}--\ref{fig:maps5}.

Assuming that the line-of-sight velocities are given by disk rotation, we can derive the maximum rotation velocity ($\vmax$) that can be compared to the velocity dispersion to study how EMPGs are dynamically supported.
We model the disk rotation using a software named GalPak$^\mathrm{3D}$ \citep[][]{Bouche+15}.
We follow the method described in Paper IX to prepare the input data cube.
Because GalPak$^\mathrm{3D}$ requires the input data cube to have the same pixel scale on the x- and y-axis, we interpolate the data cube on the y-axis to have a pixel scale of $\sim0.217''/\mathrm{pix}$.
We also correct the wavelength by the slit-width effect.
We only use the spaxels within the apertures that are determined by the source detection procedure in Section \ref{multi_clump} and mask out the rest of the 3D data cube.
We mask out the spaxels with S/N(H$\alpha$)$<3$.
We calculate the 84th percentile of $\sigma$ for the spaxels left and mask out the spaxels with $\sigma$ larger than the 84 percentile because the spaxles with large $\sigma$ could be strongly turbulent \citep[][]{Egorov+21}.
We then fit the 3D data cube by a thick disk model.
Specifically, we choose a disk model with the disk height equals to one third of the effective radius.
The H$\alpha$ surface brightness is an exponential function of radius, i.e., Sersic profile with a Sersic index of one.
We choose the arctan rotation curve with two parameters, maximum rotation velocity ($\vmax$) and turn-over radius ($r_\mathrm{v}$).
% \revise{TODO: while Sersic and arctan rotation curve}
The disk model consists of 10 parameters in total that are x- and y-coordinates of the center, total flux, effective radius ($r_\mathrm{e}$), $r_\mathrm{v}$, inclination ($i$), position angle, systemic velocity, $\vmax$, and intrinsic dispersion ($\sigma_0$).
The intrinsic dispersion is free from the broadening given by instrument, the local isotropic velocity dispersion driven by disk self-gravity, and the mixture of the line-of-sight velocities due to the disk thickness \cite[][]{Bouche+15}.
We fit the disk model to all the 33 EMPGs.
For EMPGs with compact sizes, we find the best-fit $r_\mathrm{e}$ is below the seeing size, in which case the parameters are not well constrained (see Section 4.3 in \citealt{Bouche+15}).
We thus only report 9 spatially resolved EMPGs with reliable $v_\mathrm{max}$ and $\sigma_0$ measurements that are shown in Table \ref{tab:res}.
The measurements of $\vmax$ and $\sigma_0$ for the 6 EMPGs given by Paper IX are also included.
We obtain average values of $\vmax=15~\mathrm{km~s^{-1}}$ and $\sigma_0=27~\mathrm{km~s^{-1}}$, with standard errors of $3~\mathrm{km~s^{-1}}$ and $10~\mathrm{km~s^{-1}}$, respectively.
We confirm that all of the 15($=9+6$) EMPGs have $\voversigma<1$, indicating that they may be dispersion dominated \citep{FS09}.

We show $\voversigma$ as a function of $M_*$ and metallicity in the top panels of Figure \ref{fig:relations}.
We also include the $\voversigma$ values of dwarf galaxies investigated by the DYNAMO survey \citep[][]{Green+14} and the SH$\alpha$DE survey \citep[][]{Barat+20} as comparisons.
Their results are also derived from the ionized gas.
We show that EMPGs have low $\voversigma$ that is smaller than those of more massive galaxies and comparable to those of low-mass galaxies.

\begin{deluxetable*}{lccclllll}
    \tablecaption{Summary of Galaxy Properties}
    \tablehead{
        \colhead{ID} &
        \colhead{$\log{M_*}$} & \colhead{$\log{\mathrm{SFR}}$} & \colhead{$\logOH$} &
        \colhead{$\vmax$} & \colhead{$\sigma_0$} & \colhead{$\voversigma$} &
        \colhead{$\fgas$} & \colhead{Q} \\
        \colhead{} &
        \colhead{($\log{\mathrm{M_\odot}}$)} & \colhead{($\log{\mathrm{M_\odot~yr^{-1}}}$)} & 
        \colhead{} & \colhead{(km s$^{-1}$)} & \colhead{(km s$^{-1}$)} & \colhead{} &
        \colhead{} & \colhead{}\\
        \colhead{(1)} &
        \colhead{(2)} & \colhead{(3)} & 
        \colhead{(4)} & \colhead{(5)} & \colhead{(6)} & \colhead{(7)} &
        \colhead{(8)} & \colhead{(9)}
    }
    \startdata
J0036+0052 & 7.22$\pm$0.87 & -1.57 & 7.83 & 12.8$\pm$1.9 & 24.6$\pm$1.9 & 0.52$\pm$0.09 & $0.72_{-0.29}^{+0.27}$ & $3.76_{-1.20}^{+1.25}$ \\
J0057-0941-\#1 & 6.50$\pm$0.87 & -2.67 & 7.33 & \nodata & \nodata & \nodata & $0.61_{-0.29}^{+0.38}$ & \nodata \\
J0057-0941-\#2 & 6.87$\pm$0.87 & -2.63 & 7.56 & \nodata & \nodata & \nodata & $0.45_{-0.26}^{+0.52}$ & \nodata \\
J0125+0759-\#1 & 6.24$\pm$0.87 & -1.63 & 7.47 & \nodata & \nodata & \nodata & $0.93_{-0.13}^{+0.06}$ & \nodata \\
J0125+0759-\#2 & 5.33$\pm$0.87 & -2.60 & 7.45 & \nodata & \nodata & \nodata & $0.96_{-0.09}^{+0.04}$ & \nodata \\
J0125+0759-\#3 & 5.81$\pm$0.87 & -2.13 & 7.54 & \nodata & \nodata & \nodata & $0.94_{-0.11}^{+0.05}$ & \nodata \\
J0159+0751 & 6.69$^{\color{blue}1}$ & -0.68 & 7.79 & \nodata & \nodata & \nodata & $0.99_{-0.03}^{+0.01}$ & \nodata \\
J0228-0210 & 6.87$\pm$0.87 & -1.96 & 7.12 & \nodata & \nodata & \nodata & $0.81_{-0.26}^{+0.19}$ & \nodata \\
SBS0335-052E & 7.60$\pm$0.10$^{\color{blue}2}$ & -0.61 & 7.23 & 19.7$\pm$2.9$^\mathrm{\dag}$ & 27.1$\pm$0.3$^\mathrm{\dag}$ & 0.73$\pm$0.12$^\mathrm{\dag}$ & $0.74_{-0.51}^{+0.17\mathrm{\dag}}$ & $2.62_{-0.63}^{+1.41\mathrm{\dag}}$ \\
J0811+4730 & 6.24$\pm$0.33$^{\color{blue}3}$ & -2.78 & 7.12 & \nodata & \nodata & \nodata & $0.78_{-0.27}^{+0.22}$ & \nodata \\
HS0822+3542 & 4.60$\pm$0.30$^{\color{blue}4}$ & -2.69 & 7.36 & 4.5$\pm$2.9$^\mathrm{\dag}$ & 16.6$\pm$0.5$^\mathrm{\dag}$ & 0.27$\pm$0.17$^\mathrm{\dag}$ & $0.97_{-0.07}^{+0.03\mathrm{\dag}}$ & $5.40_{-3.36}^{+3.38\mathrm{\dag}}$ \\
J0840+4707 & 7.95$\pm$0.87 & -0.11 & 7.64 & 45.0$\pm$2.6 & 51.3$\pm$1.9 & 0.88$\pm$0.06 & $0.89_{-0.18}^{+0.11}$ & $1.80_{-0.23}^{+0.35}$ \\
I Zw 18-NW & 7.24$^{\color{blue}5}$ & -1.99 & 7.35 & 6.6$\pm$2.9$^\mathrm{\dag}$ & 22.9$\pm$0.4$^\mathrm{\dag}$ & 0.29$\pm$0.13$^\mathrm{\dag}$ & $0.42_{-0.32}^{+0.37\mathrm{\dag}}$ & $11.61_{-6.85}^{+6.45\mathrm{\dag}}$ \\
I Zw 18-SE & 6.40$^{\color{blue}5}$ & -2.63 & 7.04 & \nodata & \nodata & \nodata & $0.51_{-0.28}^{+0.47}$ & \nodata \\
J0935-0115 & 6.16$\pm$0.14$^{\color{blue}6}$ & -1.55 & 7.13 & \nodata & \nodata & \nodata & $0.96_{-0.09}^{+0.04}$ & \nodata \\
J0943+3326 & 5.56$\pm$0.30$^{\color{blue}7}$ & -3.66 & 7.09 & \nodata & \nodata & \nodata & $0.52_{-0.28}^{+0.47}$ & \nodata \\
DDO 68-\#2 & 4.03$\pm$0.88 & -3.44 & 6.95 & 10.0$\pm$2.1 & 18.8$\pm$2.0 & 0.53$\pm$0.13 & $0.99_{-0.03}^{+0.01}$ & $2.71_{-0.64}^{+0.65}$ \\
DDO 68-\#3 & 4.04$\pm$0.88 & -3.61 & 6.86 & 1.5$\pm$2.1 & 22.4$\pm$2.0 & 0.07$\pm$0.10 & $0.98_{-0.04}^{+0.02}$ & $21.05_{-29.45}^{+29.46}$ \\
J1016+3754 & 6.59$\pm$0.87 & -2.03 & 7.88 & \nodata & \nodata & \nodata & $0.70_{-0.29}^{+0.29}$ & \nodata \\
Leo P & 5.75$\pm$0.09$^{\color{blue}8}$ & -4.28 & 7.02 & 2.7$\pm$2.2 & 15.9$\pm$1.9 & 0.17$\pm$0.14 & $0.12_{-0.08}^{+0.75}$ & $67.92_{-75.03}^{+55.44}$ \\
J1044+0353 & 6.04$\pm$0.07$^{\color{blue}6}$ & -1.04 & 7.44 & 14.8$\pm$4.2$^\mathrm{\dag}$ & 31.4$\pm$0.3$^\mathrm{\dag}$ & 0.47$\pm$0.14$^\mathrm{\dag}$ & $0.98_{-0.09}^{+0.01\mathrm{\dag}}$ & $3.07_{-0.92}^{+0.96\mathrm{\dag}}$ \\
J1234+3901 & 7.13$\pm$0.30$^{\color{blue}3}$ & -0.65 & 7.08 & \nodata & \nodata & \nodata & $0.98_{-0.05}^{+0.02}$ & \nodata \\
J1418+2102 & 6.63$\pm$0.15$^{\color{blue}9}$ & -1.38 & 7.52 & 22.6$\pm$2.1 & 23.9$\pm$1.9 & 0.95$\pm$0.12 & $0.90_{-0.18}^{+0.10}$ & $1.66_{-0.26}^{+0.36}$ \\
J1423+2257 & 7.63$\pm$0.87 & -0.66 & 7.90 & 34.6$\pm$1.9 & 43.9$\pm$1.9 & 0.79$\pm$0.06 & $0.84_{-0.23}^{+0.15}$ & $2.13_{-0.36}^{+0.51}$ \\
J1452+0241 & 4.18$\pm$0.14$^{\color{blue}6}$ & -2.69 & 7.49 & 1.1$\pm$2.0 & 17.5$\pm$1.9 & 0.06$\pm$0.11 & $0.99_{-0.01}^{+0.00}$ & $22.04_{-38.40}^{+38.40}$ \\
J1631+4426 & 5.89$\pm$0.10$^{\color{blue}10}$ & -1.77 & 7.24 & 7.9$\pm$1.8$^\mathrm{\dag}$ & 25.6$\pm$0.3$^\mathrm{\dag}$ & 0.31$\pm$0.08$^\mathrm{\dag}$ & $0.91_{-0.34}^{+0.06\mathrm{\dag}}$ & $5.01_{-1.27}^{+2.11\mathrm{\dag}}$ \\
J1702+2120 & 7.46$\pm$0.87 & -1.09 & 7.76 & 11.5$\pm$2.0 & 33.4$\pm$1.9 & 0.34$\pm$0.06 & $0.81_{-0.25}^{+0.19}$ & $5.09_{-1.33}^{+1.59}$ \\
J2104-0035-\#1 & 4.69$\pm$0.87 & -2.66 & 7.20 & \nodata & \nodata & \nodata & $0.98_{-0.04}^{+0.02}$ & \nodata \\
J2104-0035-\#2 & 6.01$\pm$0.87 & -3.63 & 6.97 & \nodata & \nodata & \nodata & $0.36_{-0.22}^{+0.60}$ & \nodata \\
J2115-1734-\#1 & 6.56$\pm$0.02$^{\color{blue}10}$ & -0.95 & 7.76 & 23.4$\pm$2.0 & 29.3$\pm$1.9 & 0.80$\pm$0.09 & $0.97_{-0.07}^{+0.03}$ & $1.83_{-0.20}^{+0.23}$ \\
J2115-1734-\#2 & 7.17$\pm$0.87 & -1.62 & 7.83 & \nodata & \nodata & \nodata & $0.94_{-0.18}^{+0.04\mathrm{\dag}}$ & $1.88_{-0.71}^{+0.79\mathrm{\dag}}$ \\
J2136+0414 & 6.53$\pm$0.87 & -1.72 & 7.49 & \nodata & \nodata & \nodata & $0.89_{-0.18}^{+0.10}$ & \nodata \\
J2302+0049 & 7.22$\pm$0.87 & -0.94 & 7.62 & \nodata & \nodata & \nodata & $0.90_{-0.17}^{+0.10}$ & \nodata
    \enddata
    \tablecomments{Columns: (1) ID. (2) Stellar masses. References: $^{\color{blue}1}$\cite{Izotov+17}, $^{\color{blue}2}$\cite{Pustilnik+04}, $^{\color{blue}3}$\cite{Izotov+12}, $^{\color{blue}4}$\cite{Annibali+13}, $^{\color{blue}5}$Paper IX, $^{\color{blue}6}$\cite{Xu+22}, $^{\color{blue}7}$\cite{Hirschauer+16}, $^{\color{blue}8}$\cite{McQuinn+15}, $^{\color{blue}9}$\cite{Filho+13}, $^{\color{blue}10}$\cite{Kojima+20} 
    (3) SFR calculated from the dust attenuation corrected H$\alpha$ line fluxes. The relative uncertainties are smaller than one percent in linear scale. (4) Metallicity estimated from the R3 index of the integrated {\sc [Oiii]} and H$\beta$ flux. (5)--(7) Maximum rotation velocity, intrinsic velocity dispersion, and their ratio given by the best-fit rotation disk model. (8) Gas mass fraction. (9) Global Toomre-Q parameter.}
    \tablenotetext{\dag}{Paper IX}
    \label{tab:res}
\end{deluxetable*}

\begin{figure}[ht!]
    \centering
    \includegraphics[width=\linewidth]{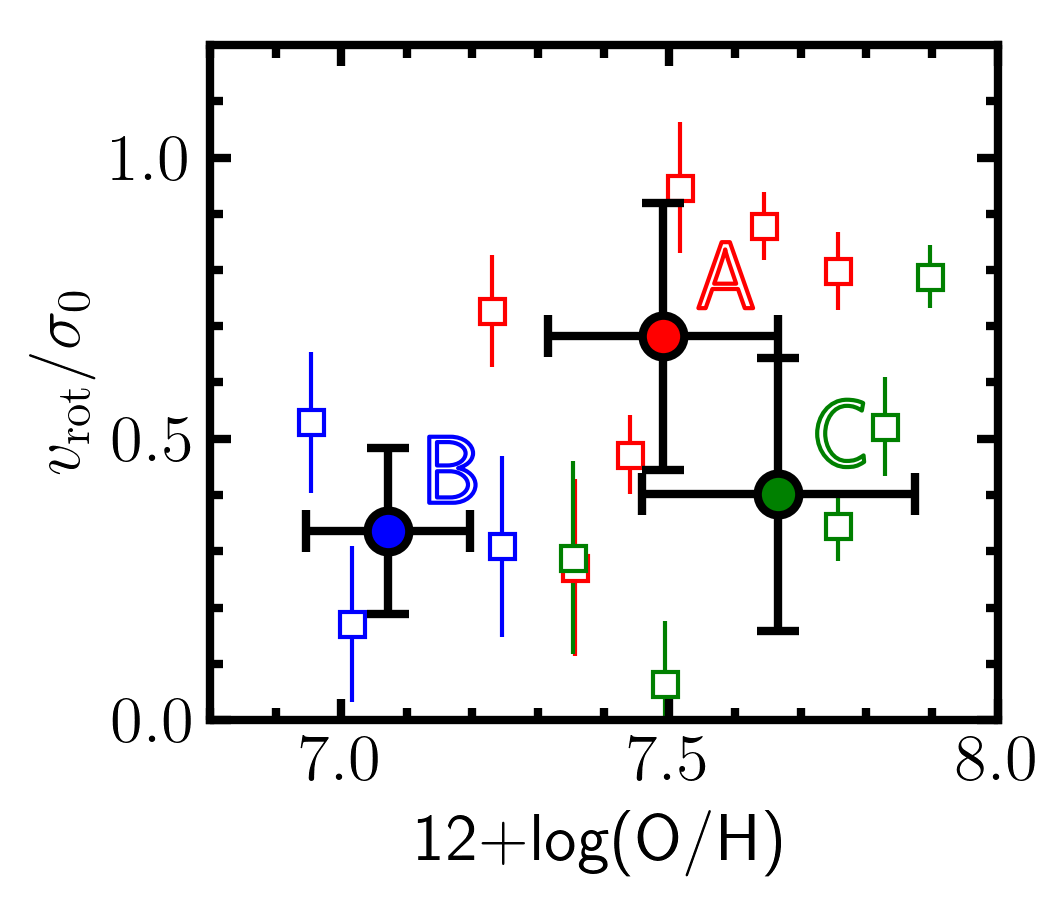}
    \caption{Same as the top-right panel of Figure \ref{fig:relations} but color-coded by the categories given by Kimihiko Nakajima et al. (in prep.). The y-axis is changed to linear scale. The category A EMPGs have surrounding metal enriched regions that may come from older stellar population while category B EMPGs only show metal poor regions given by the recent star formation. The category C EMPGs are possibly in a transition stage.}
    \label{fig:v_sigma_Nak}
\end{figure}

\section{Discussion}
\label{discuss}

\subsection{$\voversigma$ scaling relations for EMPGs}

Galaxies that reside in low-mass DM halos with shallow gravitational potential are more likely to have low $\voversigma$.
Our 15 EMPGs clearly fall into the regime of dispersion dominated galaxies having an average $\voversigma=0.48\pm0.28$.
\revise{As a comparison, dwarf galaxies ($10^6<M_*<10^9~\mathrm{M_\odot}$) studied in \cite{Barat+20} have an average $\voversigma$ of $0.84$. 
\cite{delosReyes+23} also obtain low $\voversigma\lesssim2$ based on the stellar kinematics of local dwarf galaxies ($10^7<M_*<10^9~\mathrm{M_\odot}$) and suggest a positive correlation that is independent of environmental effect.
We note that EMPGs are selected with low-metallicity and high specific star formation rate ($\log(\mathrm{sSFR/Gyr^{-1}})\sim1-3$; \citealt{Kojima+20}) compared to the more uniform samples used by \cite{Barat+20} and \cite{delosReyes+23}.}
It is curious whether the correlation between $\voversigma$ and $M_*$ are still present for \revise{EMPGs that undergo recent star formation}.
We divide our sample into two stellar mass bins of $M_* < 10^6~\mathrm{M_\odot}$ and $>10^6~\mathrm{M_\odot}$.
We obtain $\voversigma=0.24\pm0.16$ and $0.64\pm0.22$, respectively.
As shown in Figure \ref{fig:relations}, EMPGs with $M_* < 10^6~\mathrm{M_\odot}$ have smaller $\voversigma$ than those with $M_*>10^6~\mathrm{M_\odot}$ with at least $1\sigma$ significance.
For the 6 galaxies (HS 0822+3542, \revise{DDO 68-\#2, -\#3, Leo P}, J1452+0241, and J1631+4426) with $M_* < 10^6~M_\odot$, we find only weak velocity gradient in the velocity maps shown in Figure \ref{fig:maps}, consistent with the conclusion that they may not present any rotation.
\st{For local group dwarf spheroidals with $10^{3.5}~\mathrm{M_\odot}<M_*<10^8~\mathrm{M_\odot}$, Wheeler+17 study the stellar kinematics and find no clear correlation between $\voversigma$ and $M_*$.
delosReyes+23 also obtain low $\voversigma$ based on the stellar kinematics of local dwarf galaxies and suggest a positive correlation that is independent of environmental effect.
We note that our EMPGs are different from the dwarf galaxies investigated by Wheeler+17 and delosReyes+23 as EMPGs are young systems that probably undergo the first star formation activity.}

In this study we present a reasonably large sample of EMPGs to investigate the correlation between $\voversigma$ and metallicity.
We find a positive correlation as shown in the top right panel of Figure \ref{fig:relations} with a Pearson coefficient of 0.50 ($p=0.06$).
The small $\voversigma$ for the EMPGs with the smallest metallicity probably suggests that they are experiencing the first star formation activity due to gas inflow.
To test this hypothesis, we include the classification of EMPGs based on the spatial distribution of metallicity that is taken from a companion paper.
Kimihiko Nakajima et al. (in prep.) divide our EMPGs into four categories: Category A with a metal poor region in the center and relatively metal enriched around it, Category B with only metal poor region, Category C in the transitioned phase, and the unresolved Category D.
One scenario to explain the metallicity distribution of Category A EMPGs is that cold gas inflow accretes directly into the center of EMPGs (e.g., \citealt{SA14b}, Kimihiko Nakajima et al. in prep.).
In Figure \ref{fig:v_sigma_Nak}, we show $\voversigma$ as a function of metallicity with each EMPG color coded by the categories.
Surprisingly, EMPGs of Category A have $\voversigma\sim0.68\pm0.10$ on average which is larger than that $\voversigma\sim0.34\pm0.08$ for EMPGs of Category B, at 1$\sigma$ significance. 
In a Category A EMPG, the metal poorest region is surrounded by the metal enriched regions that may represent an older stellar population.
The older stellar population may have already formed a rotation disk which is destroyed by the latest gas inflow.
For a Category B EMPG, neither older stellar population nor rotation are detected.  
The EMPGs in Category B may undergo its first chemical evolutionary event that could be triggered by gas accretion.

Whether stable rotation disk can build up in low-mass star-forming galaxies like EMPGs needs to be tested with simulations \revise{(see also the discussions in Paper IX)}.
\cite{Hopkins+23} show that a sufficiently centrally-concentrated mass profile is crucial for the initial formation of a disk.
However, it is still a difficult task to resolve the profile of mass concentration \citep[e.g.,][]{Genzel+20} for \revise{compact} low-mass \st{young} galaxies by observations.
High spatial resolution with future observational facilities may help us understand how EMPGs start the recent star formation and why they appear to be dispersion dominated.
On the other hand, observations of primordial galaxies at high-$z$ (e.g., with JWST) are useful to reveal the relation between dynamics and star formation at the early stage of galaxy formation.

\begin{figure*}[ht!]
    \centering
    \plotone{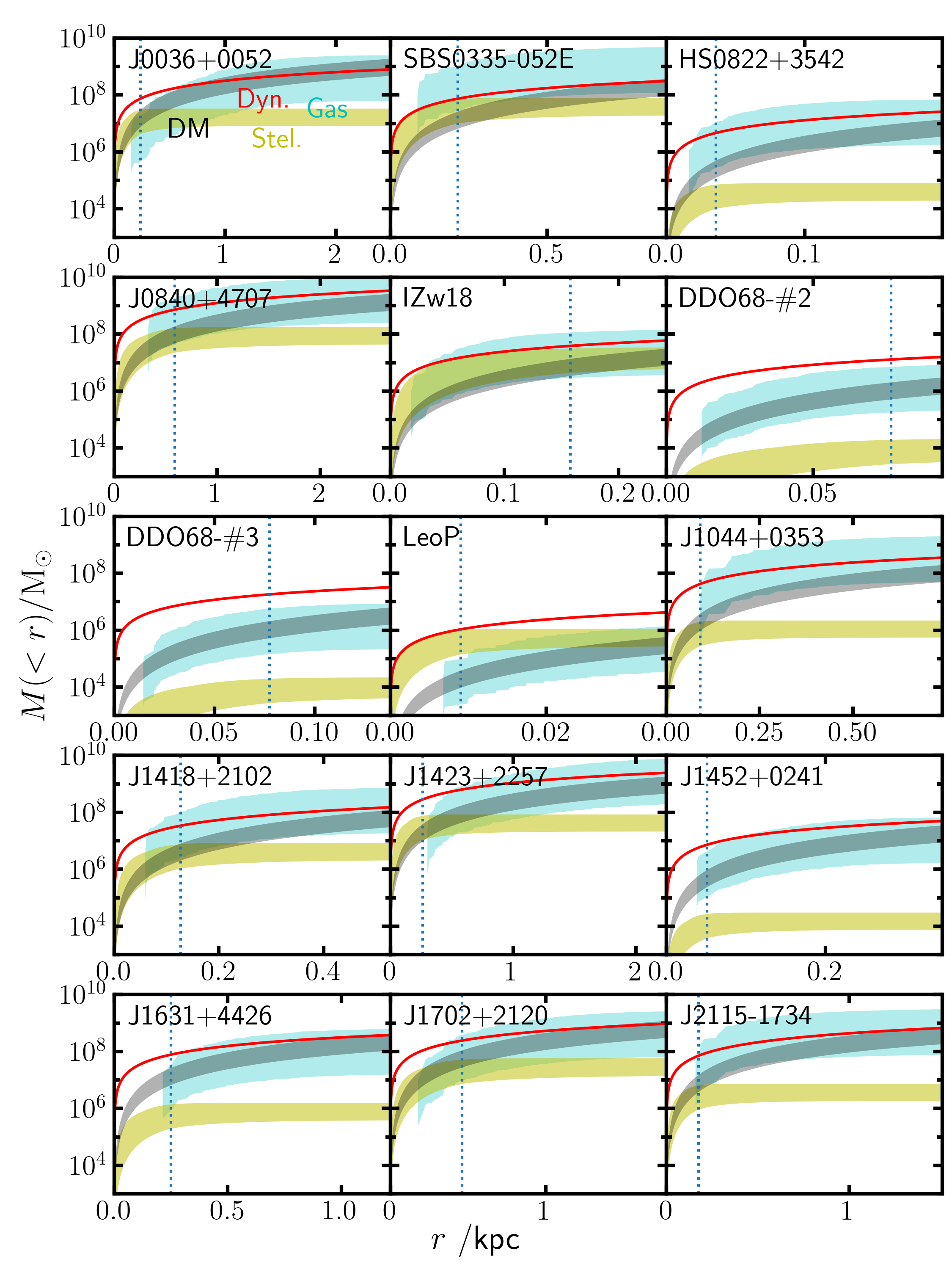}
    \caption{Enclosed mass profiles. The red, yellow, cyan, and black curves represent dynamical, stellar, gas, and DM mass profiles, respectively. The vertical dotted lines show the effective radius of H$\alpha$. The edge of the plots correspond to the outer most radii used for the kinematic analysis.}
    \label{fig:enclosed_mass}
\end{figure*}

\subsection{Mass profiles}

For the 15 EMPGs with $\vmax$ and $\sigma_0$ measurements, we can compare the radial profile of gas mass and stellar mass to that of the dynamical mass.
The dynamical mass enclosed by radius $r$ can be calculated with:
\begin{align}
    M_\mathrm{dyn}=&2.33\times10^5\left(\frac{r}{\mathrm{kpc}}\right) \notag \\
    &\left[\left(\frac{v(r)}{\mathrm{km~s^{-1}}}\right)^2+2\left(\frac{\sigma_0}{\mathrm{km~s^{-1}}}\right)^2\right]~\mathrm{M_\odot}.
\end{align}
We derive the enclosed dynamical, stellar, gas, and DM mass enclosed by radius $r$ following the procedures in Paper IX.
We follow the same procedure as Paper IX and assume a Sersic profile and Navarro–Frenk–White \citep[NFW;][]{Navarro+96} profile for the stellar mass and DM mass, respectively.
The results are plotted in Figure \ref{fig:enclosed_mass} for the 15 EMPGs, including the 6 EMPGs presented in Paper IX.
For 14 out of the 15 EMPGs, the mass profile of gas mass contribute to most of the dynamical mass, which is consistent with the large $\fgas$ we derived.
In general, we find EMPGs are likely puffly gas-rich systems supported by the random motion that may be supplied by gas inflow, stellar feedback, or galaxy-galaxy interaction.

\begin{figure*}[ht!]
    \centering
    \plotone{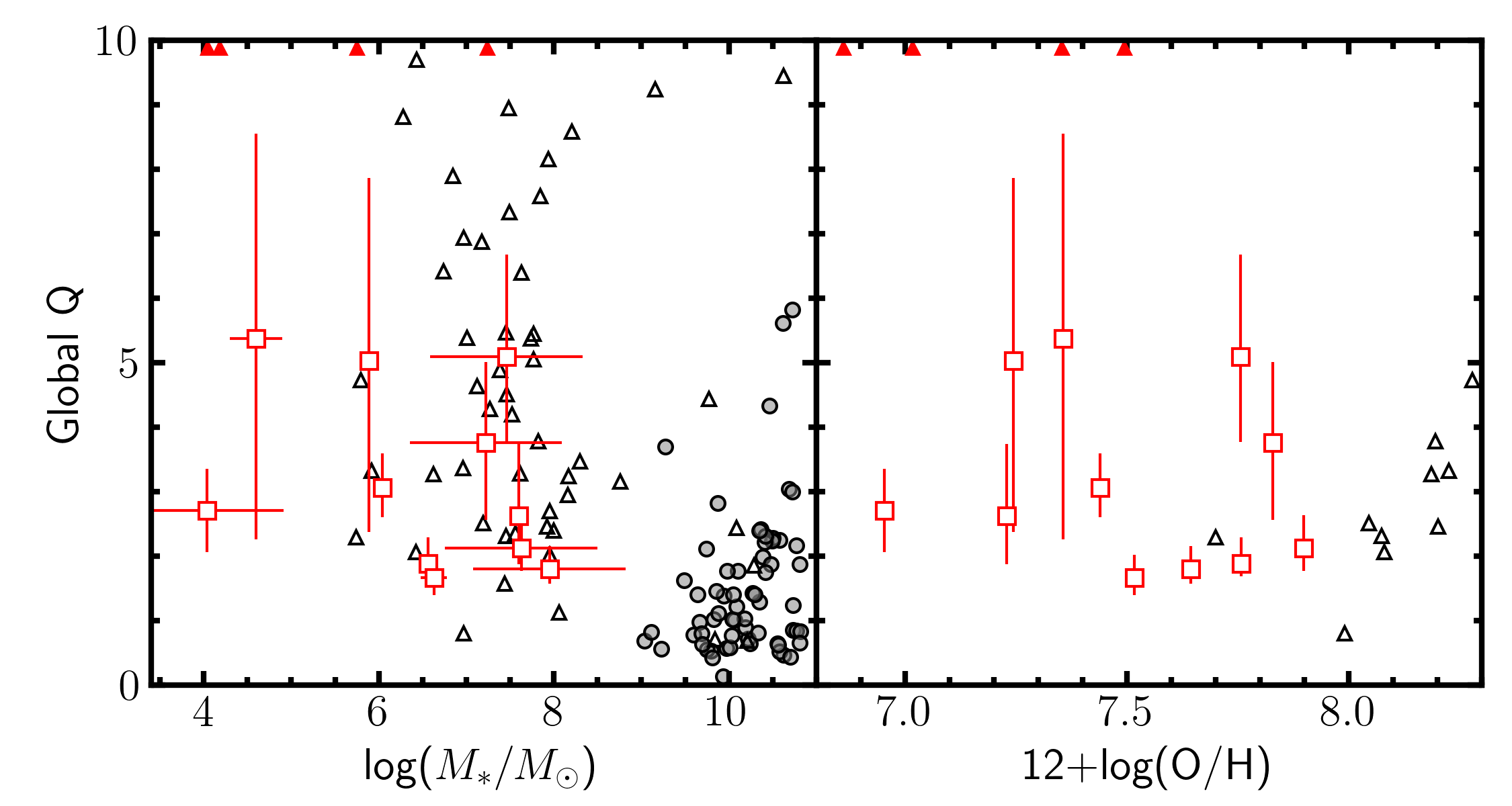}
    \caption{The global Toomre Q parameter as a function of $M_*$ (left) and metallicity (right). The red triangles indicate the four EMPGs with large global Toomre Q values that are out of the y-axis range. All of the EMPGs shown in this plot have $Q>1$.}
    \label{fig:Q}
\end{figure*}

\subsection{Toomre Q parameter}

The Toomre Q parameter is used by many kinematic studies as an indicator of the gravitational stability of disk galaxies \citep[e.g.,][]{Genzel+11}.
In general, if the Q value of a rotating disk is greater than unity (i.e., Q $> 1$), the disk is thought to be gravitationally stable. On the other hand, the disk is gravitationally unstable if Q $< 1$. 
However, it remains an open question on what observable scales the Toomre Q parameter is a reliable indicator of gravitational stability \citep[e.g.,][]{Romeo+14}.
Paper IX also suggests that it is unclear if this criterion is applicable for the EMPGs because they may not have rotating disks. 
To compare with previous kinematics studies, we calculate the average of Q within a disk that is called the global Q \citep[e.g.,][]{Aumer+10} \revise{in the same manner as Paper IX}:
\begin{equation}
    Q=\frac{\sigma_0}{\vmax}\frac{a}{\fgas},
    \label{eq:ToomreQ}
\end{equation}
where the parameter $a$ ranges from 1 to 2 depending on the gas distribution.
We assume $a=\sqrt{2}$ which corresponds to a disk with constant rotational velocity \citep[][]{Genzel+11}.
\revise{The differences of Toomre Q parameters is less than a factor of 2 if a different value of $a$ is assumed.}

We obtain a global Q of $1-70$ for 15 EMPGs with a median value of $4$.
Although all of the 15 EMPGs have global Q larger than unity suggestive of stable disk, the large global Q is inconsistent with the star-forming nature of EMPGs as pointed out in Paper IX.
In Figure \ref{fig:Q} we plot the global Q as a function of stellar mass and metallicity.
We find no clear correlation between the global Q and stellar mass or metallicity.
It is possible that EMPGs have a large variety of dynamics, or the global Q does not provide a good \revise{constraint} on the gravitational stability of EMPGs.
It is difficult to conclude whether or not EMPGs generally have stable disk without high-resolution observations \revise{for both gas} and stellar component \citep[e.g.,][]{Romeo20,Romeo+20} \revise{using next-generation facilities such as ngVLA}.

\section{Summary}
\label{sum}

We present demography of the dynamics and gas-mass fraction of 33 EMPGs with metallicities of $0.015-0.195~Z_\odot$ and low stellar masses of $10^4-10^8~\mathrm{M_\odot}$ in the local universe.
We conduct deep optical integral-field spectroscopy (IFS) for the low-mass EMPGs with the medium high resolution ($R=7500$) grism of the 8m-Subaru FOCAS IFU instrument by the EMPRESS 3D survey, and investigate H$\alpha$ emission of the EMPGs (Section \ref{data}).
Exploiting the resolution high enough for the low-mass galaxies, we derive gas dynamics with the H$\alpha$ lines by the fitting of 3-dimensional disk models. 
We obtain an average maximum rotation velocity ($\vmax$) of $15\pm3~\mathrm{km~s^{-1}}$ and an average intrinsic velocity dispersion ($\sigma_0$) of $27\pm10~\mathrm{km~s^{-1}}$ for 15 spatially resolved EMPGs out of the 33 EMPGs, and find that all of the 15 EMPGs have $\voversigma<1$ suggesting dispersion dominated systems (Section \ref{kinematics}). 
There is a clear decreasing trend of $\voversigma$ with the decreasing stellar mass and metallicity (Section \ref{discuss}). 
We derive the gas mass fraction ($f_\mathrm{gas}$) for all of the 33 EMPGs, 
and find no clear dependence on stellar mass and metallicity (Section \ref{fgas}).
Our results suggest EMPGs are gas-rich dispersion-dominated systems, whose dynamical properties likely depend on the current stellar mass and previous star formation history.
\newline
\newline
We thank the anonymous referee for constructive comments and suggestions.
We thank the staff of Subaru Telescope for their help with the observations. 
This research is based on data collected at the Subaru Telescope, which is operated by the National Astronomical Observatory of Japan (NAOJ).
We are honored and grateful for the opportunity of observing the Universe from Maunakea, which has the cultural, historical, and natural significance in Hawaii.
The Hyper Suprime-Cam (HSC) collaboration includes the astronomical communities of Japan and Taiwan, and Princeton University.
The HSC instrumentation and software were developed by the NAOJ, the Kavli Institute for the Physics and Mathematics of the Universe (Kavli IPMU), the University of Tokyo, the High Energy Accelerator Research Organization (KEK), the Academia Sinica Institute for Astronomy and Astrophysics in Taiwan (ASIAA), and Princeton University.
Based on data collected at the Subaru Telescope and retrieved from the HSC data archive system, which is operated by Subaru Telescope and Astronomy Data Center at NAOJ.
This work was supported by the joint research program of the Institute for Cosmic Ray Research (ICRR), University of Tokyo. 
Y.I., K. Nakajima, Y.H., T.K., and M. Onodera are supported by JSPS KAKENHI Grant Nos. 21J20785, 20K22373, 19J01222, 18J12840, and 21K03622, respectively.
K.H. is supported by JSPS KAKENHI Grant Nos. 20H01895, 21K13909, and 21H05447.
Y.H. is supported by JSPS KAKENHI Grant Nos. 20K14532, 21H04499, 21K03614, 22H01259, and 22KJ0157.
%M.O. is supported by JSPS KAKENHI Grant No. 21K03622.
H.Y. is supported by MEXT / JSPS KAKENHI Grant Number 21H04489 and JST FOREST Program, Grant Number JP-MJFR202Z.
J.H.K acknowledges the support from the National Research Foundation of Korea (NRF) grant, No. 2021M3F7A1084525 and No. 2020R1A2C3011091 funded by the Korea government (MSIT).
This work has been supported by the Japan Society for the Promotion of Science (JSPS) Grants-in-Aid for Scientific Research (19H05076 and 21H01128).
This work has also been supported in part by the Sumitomo Foundation Fiscal 2018 Grant for Basic Science Research Projects (180923), and the Collaboration Funding of the Institute of Statistical Mathematics ``New Development of the Studies on Galaxy Evolution with a Method of Data Science''.
The Cosmic Dawn Center is funded by the Danish National Research Foundation under grant No. 140. 
S.F. acknowledges support from the European Research Council (ERC) Consolidator Grant funding scheme (project ConTExt, grant No. 648179). 
This project has received funding from the European Union's Horizon 2020 research and innovation program under the Marie Sklodowska-Curie grant agreement No. 847523 ``INTERACTIONS''.
This work is supported by World Premier International Research Center Initiative (WPI Initiative), MEXT, Japan, as well as KAKENHI Grant-in-Aid for Scientific Research (A) (15H02064, 17H01110, 17H01114, 20H00180, and 21H04467) through Japan Society for the Promotion of Science (JSPS). 
This work has been supported in part by JSPS KAKENHI Grant Nos. JP17K05382, JP20K04024, and JP21H04499 (K. Nakajima).
This research was supported by a grant from the Hayakawa Satio Fund awarded by the Astronomical Society of Japan.
%T. Hashimoto was supported by Leading Initiative for Excellent Young Researchers, MEXT, Japan (HJH02007) and by JSPS KAKENHI Grant Number (20K22358).
JHW acknowledges support from NASA grants NNX17AG23G, 80NSSC20K0520, and 80NSSC21K1053 and NSF grants OAC-1835213 and AST-2108020.

\software{FOCAS IFU pipeline \citep{Ozaki+20}, PyRAF \citep{IRAF}, Photutils \citep{photutils},  Astropy \citep{astropy:2013,astropy:2018,astropy:2022}, GalPaK$^\mathrm{3D}$ \citep{Bouche+15}, galfit \citep{Peng+02,Peng+10}}

\bibliography{ref}{}
\bibliographystyle{aasjournal}

\end{document}